\documentclass{amsart}
\usepackage[utf8]{inputenc}
\usepackage[backend=biber, style=authoryear, natbib, maxbibnames=99]{biblatex}
\usepackage[utf8]{inputenc} 
\usepackage{graphicx}
\usepackage{nicefrac}       
\usepackage{microtype}      
\usepackage{amsmath, amsfonts,bm,amssymb,indentfirst,mathtools,bbm}
\usepackage[boxed]{algorithm2e}
\SetAlCapSkip{1em}
\usepackage{color}
\usepackage[title]{appendix}

\allowdisplaybreaks[4]

\numberwithin{figure}{section}
\numberwithin{equation}{section}

\newtheorem{theorem}{Theorem}
\newtheorem{definition}{Definition}
\newtheorem{lemma}{Lemma}
\newtheorem{remark}{Remark}

\def\cF{\mathcal{F}}
\def\cN{\mathcal{N}}
\def\cU{\mathcal{U}}
\def\trace{\text{Tr}}
\def\R{\mathbb R}

\makeatletter
\newcommand*{\rom}[1]{\expandafter\@slowromancap\romannumeral #1@}

\newcommand\niton{\mathrel{\m@th\mathpalette\canc@l\owns}}
\newcommand\canc@l[2]{{\ooalign{$\hfil#1/\mkern1mu\hfil$\crcr$#1#2$}}}
\makeatother

\usepackage{fancyhdr}
\usepackage{hyperref}
\hypersetup{colorlinks=true}

\usepackage[foot]{amsaddr}

\title[Spectral Testing Framework for Populations of Networks]{A spectral-based framework for hypothesis testing in populations of networks}

\author{Li Chen$^1$}
\author{Nathaniel Josephs$^2$}
\author{Lizhen Lin$^3$}
\author{Jie Zhou$^1$}
\author{Eric D. Kolaczyk$^2$}

\address[1]{College of Mathematics, Sichuan University, China}
\address[2]{Department of Mathematics and Statistics, Boston University, USA}
\address[3]{Department of Applied and Computational Mathematics and Statistics, The University of Notre Dame, USA}

\begin{document}

\maketitle

\begin{abstract}
    In this paper, we propose a new spectral-based approach to hypothesis testing for populations of networks.
	The primary goal is to develop a test to determine whether two given samples of networks come from the same random model or distribution.
	Our test statistic is based on the trace of the third order for a centered and scaled adjacency matrix, which we prove converges to the standard normal distribution 
	as the number of nodes tends to infinity.
	The asymptotic power guarantee of the test is also provided.
	The proper interplay between the number of networks and the number of nodes for each network is explored in characterizing the theoretical properties of the proposed testing statistics.
	Our tests are applicable to both binary and weighted networks, operate under a very general framework where the networks are allowed to be large and sparse, and can be extended to multiple-sample testing.
	We provide an extensive simulation study to demonstrate the superior performance of our test over existing methods and apply our test to three real datasets. 
\end{abstract}

\section{Introduction}\label{sec-introduction}

In this work, we consider an inference problem related to populations of networks in which each sample or data point is a network.
The statistical network analysis literature has been largely focused on proposing models and algorithms for analyzing a single network.  However, the increasing prevalence of multiple network datasets, in which the network is the fundamental data object, along with the need to extract useful scientific information from them, have motivated the demand for developing statistical methods of inference for populations of networks.

For example, in brain network data (see the COBRE data of Section \ref{sec:realdata}), one may be interested in testing whether the brain network structure from a group of individuals with schizophrenia is different from that of a group of healthy controls.
Given a collection or sample of such networks, one might also be interested in estimating some mean network feature, which could provide a notion either of averaging networks or of clustering networks into different groups \citep{mukherjee2017clustering}.
All of these cases are inference tasks for one or two samples of network objects, both of which have been recently explored in the literature.

In \citet{ginestet2017hypothesis}, the authors consider two-sample testing for networks with applications to functional neuroimaging.
Very recently, this work was extended in \citet{kolaczyk2020} through a geometric and statistical framework for inference on populations of unlabeled networks by providing a geometric characterization of the space of unlabeled networks and deriving a central limit theorem for the sample Fr\'echet mean.
Supervised and unsupervised learning such as clustering, regression, and classification for network objects have also been considered in the literature.
See, e.g., \citet{relion2019network} and \citet{josephs2020bayesian}, with the former considering network classification in neuroimaging and the latter employing Bayesian methods for classification, anomaly detection, and survival analysis.

Herein, we  focus on the problem of two-sample hypothesis testing for populations of networks. Compared to much of the work in the literature, such as \citet{ginestet2017hypothesis}, in which the number of nodes is fixed, we consider a general framework which allows both the number of nodes and the sample size (the number of networks) to grow.
In another related work, \citet{Ghoshdastidar2020} study two-sample problems from a minimax testing perspective  on testing whether two samples of binary networks of $n$ nodes are generated from the same link probability matrix against an alternative that  says the two link probability matrices are $\rho$ apart with respect to some  matrix norm. Their work focuses on the theoretical characterization of minimax separation with respect to the number of networks $m$, the number of nodes $n$, and different matrix norms.  \citet{tang2017} study whether two random dot product networks ($m=1$) defined on different vertex sets are generated from the model or not. 


Our test statistics are spectral based and not restricted to a given network structure.
We utilize the trace of the third order for a centered and scaled adjacency matrix, which is proven to converge to the standard normal distribution as the number of nodes tends to infinity.
In addition, we show that the asymptotic power tends to one as the number of nodes increases.
Since we also want to understand the limiting behavior as the sample size increases, we explore the proper interplay between the asymptotics in the number of networks and in the number of nodes for each network when characterizing the theoretical properties of our proposed testing statistics.
These statistics are conceptually simple and computational friendly and we provide an extensive simulation study under various models to demonstrate the superior performance of our test over existing methods.
In almost all the cases examined in our study, the proposed test statistics achieve the nominal rejection rate under the null and a power close to one under the alternative. We also apply our test to three real datasets, based on both weighted and binary networks. 
 
The idea of applying a spectral method based on random matrix theory to network data is a natural one, as the network data (e.g., the adjacency or Laplacian matrix) can be naturally  viewed as a random matrix.
Spectral-based hypothesis tests, in particular, have been proposed in \citet{bickel2016hypothesis} and \citet{dong2020spectral} for testing the community structure and determining the number of clusters within a single network.
A spectral-based test based on a Tracy-Widom law for hypothesis testing of populations of networks and change point detection in networks can be found in   \citet{2019arXiv191103783C} and \citet{chen2020hypothesis}.
Compared to those two works, our spectral-based test has asymptotic standard normal distribution with much faster convergence rate under the null compared to the slow convergence of a test that has a Tracy-Widom law.
Furthermore, our testing statistics require much milder conditions for the theoretical performance guarantees: an error estimate of the link probability estimates with $o_p(1)$ is needed in comparing to an error condition of $o_p(n^{-2/3})$ required in \citet{chen2020hypothesis}.
Note that although the statistic in \citet{dong2020spectral} also has an asymptotically normal law, it is limited to testing the presence of community structure in a single network versus the null Erd\"os-R\'enyi model, whereas our statistic can test the difference between arbitrary network models and can be applied to either binary or weighted networks in both two-sample and multiple-sample frameworks.

The remainder of the paper is organized as follows.
In Section \ref{sec-test}, we describe our proposed spectral-based testing statistics and derive their asymptotic null distributions as well as asymptotic power results.
We extend our test for weighted networks and multiple-sample testing in Section \ref{sec-extend}.
Results of extensive simulation studies are reported in Section \ref{sec-simulation} and analysis on three real network datasets are given in Section \ref{sec:realdata}.
We conclude in Section \ref{discussion} with a few final remarks and possible future directions of this work.


\section{A new spectral-based test for binary networks}\label{sec-test}
	
In this section,  we first propose a new spectral-based test for testing the difference between distributions of two samples of binary networks. 
Specifically, we consider two samples of networks on the same $n$ nodes with possibly different sample sizes $m_1$ and $m_2$.
We assume one observes the independent and identically distributed symmetric binary adjacency matrices  $A_1^{(1)}, \ldots, A_1^{(m_1)}$, with conditionally independent entries generated from a symmetric link probability matrix $P_1$, i.e.
    \[
        A_{1,i j}^{(k)} \sim \text{Bernoulli}(P_{1, i j}) \enskip ,
    \]
for $k = 1,2,\ldots,m_1$, $i,j = 1,2,\ldots,n$.
Similarly, one observes a second sample of  adjacency matrices $A_2^{(1)},\ldots,A_2^{(m_2)}$ with
    \[
        A_{2,i j}^{(k)} \sim \text{Bernoulli}(P_{2, i j}) \enskip ,
    \]
generated from the same model with link probability matrix $P_2$.
Our goal is to test whether the two samples of networks have the same graph structure or not, which is equivalent to testing
    \begin{equation}\label{test for two-sample}
        H_0: P_1 = P_2 ~ \text{against} ~ H_1: P_1 \neq P_2 \enskip .
    \end{equation}
    
To address this, we propose a new statistic that utilizes results from random matrix theory.
For necessary background on spectral properties of inhomogeneous networks, which are used heavily in this work, see Appendix \ref{appendix: background}.
    
\subsection{New spectral test for binary networks}\label{sec-unweighted}
    
Given two samples of networks $\{A_1^{(k)}\}_{k = 1}^{m_1}$ and $\{A_2^{(k)}\}_{k = 1}^{m_2}$ sampled from the link probability matrices $P_1$ and $P_2$, respectively, we introduce the normalized matrix with elements as follows:
    \begin{equation}\label{Z}
        Z_{i j} = \begin{cases}
            \frac{\bar{A}_{1, i j} - \bar{A}_{2, i j}}{\sqrt{n \big(\frac{1}{m_1} P_{1, i j} (1-P_{1, i j}) + \frac{1}{m_2} P_{2, i j} (1-P_{2, i j})\big)}} & \text{ if } i \neq j \\
            B_{i j } & \text{ if } i = j
        \end{cases} \enskip ,
    \end{equation}
where $\bar{A}_{u}$ is the sample average of adjacency matrices in the $u$th group, for $u=1,2$,
    \begin{equation}\label{eq:A_bar}
        \bar{A}_{u} = \frac{1}{m_u} \sum_{k = 1}^{m_u} A_u^{(k)} \enskip ,    
    \end{equation}
and $B$ is an $n \times n$ diagonal matrix with $B_{i i}$ given by i.i.d random variables such that
    \begin{equation}\label{eq:B}
        P(B_{i i} = - 1 / \sqrt{n}) = P(B_{i i} = 1 / \sqrt{n}) = 1/2 \enskip ,
    \end{equation}
for $i = 1, \ldots, n$.
    
Consider the test statistic 
\begin{equation}
    \theta = \frac{1}{\sqrt{15}} \trace(Z^3)\enskip , \label{theta}
\end{equation}
where $\trace(\cdot)$ represents the trace operator.
We have the following theorem on the asymptotic distribution of $\theta$ under the null hypothesis. 
     
\begin{theorem}\label{theorem theta unweight}
    Let $Z$ be given as in \eqref{Z}.
    Assume the sample sizes $m_1 = O(n^{\alpha_1})$ and $m_2 = O(n^{\alpha_2})$ for some $\alpha_1, \alpha_2 \in (0, 1)$.
    Then, under the null hypothesis $P_1 = P_2$, for the scaled test statistic $\theta  = \frac{1}{\sqrt{15}} \trace(Z^3)$, 
    we have
    	\begin{equation}
	        \theta  \overset{d}{\to} \cN(0,1) \quad \text{as} \quad n \to \infty \enskip .
	    \end{equation}
\end{theorem}

We defer the details of the proof to Appendix \ref{proof: theta unweight}.
However, the overview of the argument is as follows.
First, one can see that under the null hypothesis of $P_1 = P_2$, $Z$ is a Wigner matrix satisfying $\mathrm{E}(Z_{i j}) = 0$ and $\text{Var}(Z_{i j}) = 1 / n$. 
Then the remainder of the proof proceeds in three steps.
We begin by showing that the empirical spectral distribution of $Z$ converges to the semicircular law almost surely.
Next, we verify that $X = \sqrt{n}Z$ satisfies conditions $(1)$--$(3)$ of Lemma \ref{Lemma 9.2}, after which the asymptotic normality of $\theta$ follows.
Finally, the mean and the variance are obtained from \citet{dong2020spectral}.

To formalize a testing framework using $\theta$ in \eqref{theta}, we need to account for the fact that the diagonal matrix $B$ in \eqref{eq:B} is random.
We do so by employing a Monte Carlo procedure, which we describe in Algorithm \ref{alg:test}.
Our output is an empirical confidence level, which is the rejection rate based on the test statistics computed from the Monte Carlo samples of $B$.

\begin{algorithm*}[!htb]
\caption{Procedure for testing using the statistic in \protect{\eqref{theta}}. The output is an empirical significance level based on Monte Carlo test statistics, where $I(\cdot)$ is an indicator function and $\mu_{\alpha / 2}$ is the $\alpha / 2$ upper quantile of $\cN(0, 1)$.}
\label{alg:test}
\SetKwInOut{Input}{Input}
\SetKwInOut{Output}{Output}
\underline{New Spectral-Based Hypothesis Test} $\big(\{A_1^{(k)}\}_{k=1}^{m_1}, \{A_2^{(k)}\}_{k=1}^{m_2}, \alpha, Q\big)$\;\Input{Adjacency matrices $\{A_1^{(k)}\}_{k=1}^{m_1}$ and $\{A_2^{(k)}\}_{k=1}^{m_2}$ for groups 1 and 2 \\
Significance level $\alpha$ \\
Number of Monte Carlo samples $Q$}
\Output{Empirical significance level $\text{rej\_rate}$}
Compute $\bar{A}_u$ for $u = 1, 2$ using \eqref{eq:A_bar} \;
\ForPar{$q = 1,\ldots,Q$}
{
    Sample $B^{(q)}$ satisfying \eqref{eq:B} \;
    Compute $Z^{(q)}$ in \eqref{Z} using $B^{(q)}$ \;
    Compute $\theta^{(q)}$ in \eqref{theta} using $Z^{(q)}$ \;
}
$\text{rej\_rate} = \frac{1}{Q}\sum_{q=1}^{Q} I\big(|\theta^{(q)}| > \mu_{\alpha / 2} \big)$
\end{algorithm*}


\begin{remark}
    In Algorithm \ref{alg:test}, we deliberately do not output a p-value.
    For $Q = 1$, we could obtain a p-value using $2 P\big(\theta > |\theta_{obs}^{(Q=1)}|)$
    as in \citet{bickel2016hypothesis} and \citet{dong2020spectral}, where $\theta_{obs}^{(Q=1)}$ is the sample test statistic and $\theta$ follows the null distribution of the testing statistic.
    In this case, though, the p-value is implicitly conditional on $B$ and the authors' simulations reveal that the randomness of $B$ leads to highly variable p-values.
    Instead, for our test, we propose computing many $\theta_{obs}^{(q)}$ 
    in parallel to reduce the noise induced by $B$.
    The analogous p-value estimate combining these Monte Carlo test statistics would be $2 P\big(\theta_Q > |\bar{\theta}_{obs}| \big)$, where $\bar{\theta}_{obs} = \frac{1}{Q}\sum_{q=1}^Q \theta_{obs}^{(q)}$.
\end{remark}

\begin{remark}
    The rejection rate from our Monte Carlo estimator has the property that its expectation under the null is the nominal significance level:
        \[\mathrm{E}\Big(\frac{1}{Q}\sum_{q=1}^{Q} I\big(|\theta^{(q)}| > \mu_{\alpha / 2} \big)\Big) = P\big(|\theta^{(q)}| > \mu_{\alpha / 2}\big) = \alpha \enskip .\]
\end{remark}

\subsection{Test statistic based on estimated link probability matrices}

Theorem \ref{theorem theta unweight} assumes that the true link probability matrices $P_1$ and $P_2$ are known, which is not the case in practice.
Therefore, $\theta$ cannot be used directly as a test statistic.  A natural alternative is to plug in some appropriate estimates of $P_1$ and $P_2$ with the hope that the plug-in estimator for the test statistic retains asymptotic normality.

We denote the plug-in estimates of $P_1$ and $P_2$ by $\hat P_1$ and $\hat P_2$, respectively.
Then the empirical version of the normalized matrix $Z$ in \eqref{Z} can be written as 
    \begin{equation}\label{Z_hat}
    \hat Z_{i j} = \begin{cases}    
        \frac{\bar{A}_{1, i j} - \bar{A}_{2, i j}}{\sqrt{n \left(\frac{1}{m_1} \hat P_{1, i j} (1 - \hat P_{1, i j}) + \frac{1}{m_2} \hat P_{2, i j} (1 - \hat P_{2, i j})\right)}} & \text{ if } i \neq j \\
        B_{i j} & \text{ if } i = j
        \end{cases} \enskip .
    \end{equation}
The resulting test statistic is thus
    \begin{equation}\label{theta_hat}
        \hat{\theta} = \frac{1}{\sqrt{15}} \trace{(\hat{Z}^3)} \enskip ,
    \end{equation}
which has the following limiting law.
    
\begin{theorem}\label{theorem theta hat unweight}
   Under  the two-sample framework of binary networks, let $\hat Z$ be given in \eqref{Z_hat}.
   As before, assume the sample sizes $m_1 = O(n^{\alpha_1})$ and $m_2 = O(n^{\alpha_2})$ for some $\alpha_1, \alpha_2 \in (0, 1)$.
   Suppose $\hat P_1$ and $\hat P_2$ are some estimates of $P_1$ and $P_2$, respectively.
   If $\max_{i j}|\hat{P}_{u, i j} - P_{u, i j}| =  o_p(1)$ for $u = 1, 2$, then, under the null hypothesis $P_1 = P_2$, we have the following asymptotic distribution of the scaled test statistic $\hat\theta  = \frac{1}{\sqrt{15}} \trace(\hat Z^3)$:
        \[
            \hat{\theta} \overset{d}{\to} \cN(0,1) \quad \text{as} \quad n \to \infty \enskip .
        \]
\end{theorem}
    
Again, we defer the proof to Appendix \ref{proof: theta hat unweight}, which relies on rewriting
    \[
        \trace(\hat{Z})^3 = \trace(Z^3) + 3 \trace\big( Z^2 (Z \circ H)\big) + 3 \trace \big(Z (Z \circ H)^2\big) + \trace \big((Z \circ H)^3 \big) \enskip ,
    \]
where $\circ$ denotes the Hadamard product and $H$ is an $n \times n$ matrix with entries $H_{i j} = \max_{u = 1, 2} O(\hat P_{u, i j} - P_{i j}) =  o_p(1)$.
Each term in the right of this equality can be written as an element-wise sum and can be divided into two parts in terms of the corresponding subscripts, which we bound to show convergence as $n$ goes to infinity.
      
\subsection{Estimating link probability matrices}\label{estimating link}

Theorem \ref{theorem theta hat unweight} requires the sample sizes of observed networks $m_u$ to grow with $n$ at a rate of $n^{\alpha}$ for any $\alpha \in (0,1)$ and $\max_{i j}|\hat{P}_{u, i j} - P_{u, i j}| =  o_p(1)$, which are very mild conditions satisfied by many estimation methods.  

The simplest estimator  of $P_{u, ij}$ is the sample mean of all the $(i,j)$ elements in the adjacency matrices of group $u$, where $u = 1, 2$.
We refer to this spectral method based on simple averages as SPE-AVG.
It is not difficult to see that $\max_{i,j} |\hat{P}_{u,i j} - P_{u, i j}| = o_p(m_u^{-1 / 2} \log (n))$.
Intuitively, SPE-AVG requires sample sizes to be large enough to achieve good performance.
This is also confirmed empirically by our extensive simulation study in which SPE-AVG typically yields inferior performance compared to the next methods we present.

Another possible average estimator of $P_{u, i j}$ is based on the stochastic block model (SBM).
The key idea is to approximate any graph with an SBM, which, for large networks, is reasonable by Szemer\'{e}di's regularity lemma \citep{lovasz2012}.
The membership vector of nodes can be obtained by community algorithms such as the method proposed in \citet{ng2002spectral}.
After the membership vector has been estimated, we can simply approximate $P_{u, i j}$ by the sample mean of all the entries in the submatrix over all $A_{u}^{(k)}$, $k = 1, 2, \cdots, m_u$, restricted to the corresponding block consisting of the communities of $i$ and $j$.
We refer to this test method based on SBM as SPE-SBM.
Assuming the true community number is $K_u$, then the variance error satisfies $\max_{i,j} |\hat{P}_{u,i j} - P_{u, i j}| = o_p(K_u m_u^{-1 / 2} n^{- 1} \log (n))$.
It can be seen that the rate of SPE-SBM is better than that of SPE-AVG as long as $K_u < n$, which is very easy to be satisfied. However, the property may be limited by the assumption that the network topologies follow an SBM structure.

Finally, we introduce a new method of estimation based on the modified neighborhood smoothing (MNBS) proposed in \citet{zhao2019}.
The idea is to perform neighborhood smoothing to the matrix $\bar A$, which is the weighted average of $m$ networks and the smoothing procedure is applied to a shrunken neighborhood size.
This results in a better bias-variance tradeoff compared to the neighborhood smoothing (NBS) method proposed in \citet{levina} leading to a  better estimate of the link probability matrix with a smaller error. 
When the sample size $m_u$ is small compared with $n$, which is more common in practice, and satisfies $(m_u \log n)^{1/2} < n^{1/2}$, then from Lemma 9.3 in \citet{zhao2019}, the size of neighborhood is $O_p((n \log n / m_u)^{1/2})$.
Thus the estimation error of the link probability is $|\hat{P}_{u,i j} - P_{u, i j}| = O_p((m_u n \log n)^{-1/4})$.
We refer to this test method based on MNBS as SPE-MNBS. We also note that SPE-MNBS puts no structure conditions on the networks. 
Therefore, we expect SPE-MNBS to be generally applicable.
      
\subsection{Asymptotic power guarantee}

Next we consider the power of the test based on $\hat\theta$, which we summarize in the following theorem.
      
\begin{theorem}\label{theorem power unweight}
    Consider the alternative model of $P_1 \neq P_2$ under the assumptions of Theorem \ref{theorem theta unweight}.
    Let $Z''$ be an $n \times n$ matrix with zero diagonals and, for any $i \neq j$,
        \begin{align} \label{Z''}
    	    Z''_{i j} &= \frac{P_{1, i j} - P_{2, i j}}{\sqrt{n \left(\frac{1}{m_1} P_{i j} (1 - P_{i j}) + \frac{1}{m_2} P_{i j} (1 - P_{i j})\right)}} \enskip .
    	\end{align}
    Define the partition $\{1, \cdots, n\}^3 = S_{a} \cup S_{b}$, where $(i,k,l) \in S_{a}$ indicates that $Z''_{i k} Z''_{k l} Z''_{l i} \geq 0$, and $(i,k,l) \in S_{b}$, that $Z''_{i k} Z''_{k l} Z''_{l i} < 0$.  Write
     $|S_{a}| = a n^3$ and $|S_{b}| = b n^3$, with $a, b \in [0, 1]$ satisfying $a + b = 1$.
    If either of the following conditions are satisfied,
    	\begin{align*}
    	    (i) &\quad a \min_{(i, k, l) \in S_a} ({Z''}_{i k})^3 + b \min_{(i, k, l) \in S_b} ({Z''}_{i k})^3 > 0 \enskip , \\
    	    (ii) &\quad a \max_{(i, k, l) \in S_a} ({Z''}_{i k})^3 + b \max_{(i, k, l) \in S_b} ({Z''}_{i k})^3 < 0 \enskip ,
    	\end{align*}    	
    then 
    	\[
    	    \lim_{n \to \infty} P(|\hat{\theta}| > \mu_{\alpha / 2}) = 1, \ \alpha \in (0, 1) \enskip .
    	\]
\end{theorem}

The details of the proof, which are similar to those in Theorem \ref{theorem theta hat unweight}, are given in Appendix \ref{proof: power unweight}.

\begin{remark}
Note that there is a slight abuse of notation in conditions $(i)$ and $(ii)$ where the minimum operator is taken over all pairs of indices among $(i,j,k).$
   Conditions $(i)$ and $(ii)$ above characterize the minimum signal difference between $P_1$ and $P_2$ required for Theorem \ref{theorem power unweight} to hold, which implies that the power is asymptotically one when either of the sets $S_a$ or $S_b$ is large enough.    
\end{remark}

\section{Extending our test to other settings}\label{sec-extend}

In this section, we extend our test to be used for weighted networks, as well as for multiple samples in a manner analogous to one-way analysis of variance (ANOVA).

\subsection{Extension to weighted networks}\label{sec-weighted}
    
We now consider a more general framework that focuses on weighted networks.
Let  $F_1 = \{F_{1,i j}\}$ and $F_2 = \{F_{2,i j}\}$ for $i, j = 1, \ldots, n$ be two sequences of distributions defined on bounded intervals and specified by some parameters.
Let $A_1^{(1)}, \ldots, A_1^{(m_1)} \overset{i.i.d}{\sim} F_1$ and $A_2^{(1)},\ldots, A_2^{(m_2)} \overset{i.i.d}{\sim} F_2$ be symmetric weighted adjacency matrices for networks that are undirected and without self-loops, i.e. $\ A_{u,i i}^{(k)} = 0$ for $u = 1, 2, \ i = 1, \ldots, n$, and $k = 1, \ldots, m_u$.  
Let $\Sigma_{u}$ denote an $n \times n$ matrix in which the $(i, j)$ element is the variance of $A_{u, i j}^{(k)}$ for $k = 1, 2, \ldots, m_u$.
Note that its diagonal elements are 0 since $A_{u, i i}^{(k)} = 0$.
Finally, let $\hat{\Sigma}_{u,i j}$ be an estimate of ${\Sigma}_{u,i j}$.

Our approach for weighted networks is to replace $P_{u, i j} (1 - {P}_{u, i j})$ in \eqref{Z} and $\hat P_{u, i j} (1 - \hat{P}_{u, i j})$ in \eqref{Z_hat} by ${\Sigma}_{u,i j}$ and $\hat{\Sigma}_{u,i j}$, respectively.
Just as in Section \ref{estimating link}, estimates $\hat{\Sigma}_{u,i j}$ can also be obtained using various methods, which will be discussed later.  For simplicity, we use the same notation as in Section \ref{estimating link}.
    
For the weighted case, the testing problem in \eqref{test for two-sample} is equivalent to
    \begin{equation}\label{test-weighted}
        H_0: F_1 = F_2 ~ \text{against} ~ H_1: F_1 \neq F_2 \enskip .
    \end{equation}
We define the normalized matrix $Z$ as
    \begin{equation}\label{Z weight}
        Z_{i j} = \begin{cases}
            \frac{\bar{A}_{1, i j} - \bar{A}_{2, i j}}{\sqrt{n \big(\frac{1}{m_1} \Sigma_{1, i j} + \frac{1}{m_2} \Sigma_{2, i j}\big)}} & \text{ if } i \neq j \\
            B_{i j } & \text{ if } i = j
        \end{cases} \enskip .
    \end{equation}   
Then the asymptotic distribution of  $\theta = \frac{1}{\sqrt{15}} \trace(Z^3)$ follows a standard normal distribution under the null hypothesis, as stated in the following theorem.
    
\begin{theorem}\label{theorem theta}    
    Under the two-sample framework of weighted networks, let $Z$ be given in \eqref{Z weight}. Assume sample sizes $m_1 = O(n^{\alpha_1})$ and $m_2 = O(n^{\alpha_2})$ for some $\alpha_1, \alpha_2 \in (0,1)$.
    Then, under the null hypothesis $F_1 = F_2$, for scaled test statistic $\theta = \frac{1}{\sqrt{15}} \trace(Z^3)$, we have
    	\begin{equation}
    	    \theta  \overset{d}{\to} \cN(0,1) \quad \text{as} \quad n \to \infty \enskip .
    	\end{equation}
    \end{theorem}

The proof is omitted since it is similar to that of Theorem \ref{theorem theta unweight}.

\begin{remark}
    Although the two-sample testing framework for  binary networks is a special case of that of \eqref{test-weighted}, we discuss the two cases separately.
    In the binary case, our test statistic is obtained by plugging in an estimate of the link probability  matrix $P$, while our test statistic for the weighted networks requires a plug-in estimate of the variance of each edge weight.
    Hence the estimation methods differ for these two cases.
\end{remark}
    
For practical application, the variance matrices $\Sigma_1$ and $\Sigma_2$ need to be estimated, with some conditions assumed to ensure that the asymptotic normality of the new test statistic still holds.
For $\hat\Sigma_1$ and $\hat\Sigma_2$, the plug-in estimates of $\Sigma_1$ and $\Sigma_2$, respectively, the empirical normalized matrix of $Z$ in \eqref{Z weight} can be written with entries as 
    \begin{equation}\label{Z_hat weight}
        \hat Z_{i j} = \begin{cases}
            \frac{\bar{A}_{1, i j} - \bar{A}_{2, i j}}{\sqrt{n \left(\frac{1}{m_1} \hat \Sigma_{1, i j} + \frac{1}{m_2} \hat \Sigma_{2, i j} \right)}} & \text{ if } i \neq j\\
        B_{i j } & \text{ if } i = j
        \end{cases} \enskip .
    \end{equation}
Therefore, our test statistic is  
    \begin{equation}\label{theta_hat unweight}
        \hat{\theta} = \frac{1}{\sqrt{15}} \trace{(\hat{Z}^3)} \enskip .
    \end{equation}
Then we have the following limiting law.    
\begin{theorem}\label{theorem theta hat}
    In the two-sample framework of weighted networks, let $\hat Z$ be given in \eqref{Z_hat weight}.
    Assume the sample sizes $m_1 = O(n^{\alpha_1})$ and $ m_2 = O(n^{\alpha_2})$ for some $\alpha_1, \alpha_2 \in (0,1)$. Suppose $\hat \Sigma_1$ and $\hat \Sigma_2$ are some estimates of $\Sigma_1$ and $\Sigma_2$, respectively.
    If $\max_{i j}|\hat{\Sigma}_{u, i j} - \Sigma_{u, i j}| =  o_p(1)$, $u = 1, 2$, then under the null hypothesis $F_1 = F_2$, we have the following asymptotic distribution of the scaled test statistic $\hat \theta =  \frac{1}{\sqrt{15}} \trace{(\hat{Z}^3)}$:
        \[
            \hat{\theta} \overset{d}{\to} \cN(0,1) \quad \text{as} \quad n \to \infty \enskip .
        \]
\end{theorem}

The proof is similar to that of Theorem \ref{theorem theta hat unweight}, so we only include the key differences in Appendix \ref{proof: theta hat}, as the remainder of the proof can be completed straightforwardly.
    
We consider two estimates of $\Sigma_{u, i j}$.
The first is obtained simply as the sample variance of each element over all adjacency matrices in the same group. For convenience, we still refer to this method as SPE-AVG. Then we have 
\begin{equation} \label{error}
    \max_{i, j}|\hat{\Sigma}_{u, i j} - \Sigma_{u, i j}| = O_p(m_u^{-1/2}) \enskip .
\end{equation}
The  details of the proof can be found in Appendix \ref{proof: error}. The order of the error is the same as the binary case which implies that SPE-AVG is suitable for large sample-size cases.

The second estimate of $\Sigma_{u, i j}$ is obtained similarly as SPE-SBM for unweighted networks: assume each network comes from an SBM, approximate the community membership vector, and compute the sample covariance within each community as the sample variance of the nodes corresponding to that community block (rather than the sample mean). Again, we refer to this method as SPE-SBM as in the binary case. With a similar argument as the proof in the Appendix \ref{proof: error}, one has $\max_{i, j}|\hat{\Sigma}_{u, i j} - \Sigma_{u, i j}| = O_p(K_u m_u^{-1/2} n^{-1})$. Therefore, the error condition in Theorem \ref{theorem theta hat} is  satisfied as long as $K_u < n$, which  should hold for most cases.
    
The power of the test for weighted networks is presented in the following theorem.
    
\begin{theorem}\label{theorem power}
    Under the assumptions of Theorem \ref{theorem theta} and the alternative model $F_1 \neq F_2$, let $Z''$ be an $n \times n$ matrix with zero diagonals and for any $i \neq j$,
    	\begin{align*}
        	Z''_{i j} &= \frac{P_{1, i j} - P_{2, i j}}{\sqrt{n \left(\frac{1}{m_1} \Sigma_{1,i j} + \frac{1}{m_2} \Sigma_{2,i j}\right)}} \enskip ,
    	\end{align*}
    Define $S_a$ and $S_b$ as in Theorem~\ref{theorem power unweight} based on the above $Z''$.
    If either of the following conditions are satisfied,
    	\begin{align*}
    	    (i) &\quad a \min_{(i, k, l) \in S_a} ({Z''}_{i k})^3 + b  \min_{(i, k, l) \in S_b} ({Z''}_{i k})^3 > 0 \enskip , \\
    	    (ii) &\quad a  \max_{(i, k, l)\in S_a} ({Z''}_{i k})^3 + b \max_{(i, k, l) \in S_b} ({Z''}_{i k})^3 < 0 \enskip ,
    	\end{align*}    	
    then
        \[
            \lim_{n \to \infty} P(|\hat{\theta}| > \mu_{\alpha / 2}) = 1, \ \alpha \in (0, 1) \enskip .
        \]
\end{theorem}
    
Again, the proof is omitted as it is similar to that of Theorem \ref{theorem power unweight}.
    
\subsection{Extension to multiple-sample testing}\label{sec-anova}

Finally, we consider the case when $S$ groups are present and $S > 2$.
Assume one observes the symmetric binary adjacency matrices  $A_s^{(1)}, \ldots, A_s^{(m_s)}$ that are generated from a symmetric link probability matrix $P_s$, i.e.
    \[
        A_{s,i j}^{(k)} \sim \text{Bernoulli}(P_{s, i j}) \enskip ,
    \]
for $s = 1, \ldots, S$, $k = 1,\ldots,m_s$, and $i,j = 1,\ldots,n$.
Our goal is to test whether there are any differences in the distributions of the $S$ groups, which is equivalent to testing
    \begin{equation}
        H_0: P_1 = P_2 = \cdots = P_S ~ \text{ against } ~ H_1: P_s \text{ are not all equal} \enskip .
    \end{equation}
This is analogous to one-way ANOVA.

We define the pairwise normalized matrices with elements as follows:
    \begin{equation}\label{anova}
        Z^{(s)}_{i j} = \begin{cases}
            \frac{\bar{A}_{s, i j} - \bar{A}_{i j}}{\sqrt{n \Big(\big(\frac{1}{m_s} - \frac{2}{m}\big) P_{s, i j} (1-P_{s, i j}) + \frac{1}{m^2} \sum_{s=1}^S m_s P_{s, i j} (1-P_{s, i j})}\Big)} & \text{ if } i \neq j \\
            B_{i j } & \text{ if } i = j
        \end{cases} \enskip ,
    \end{equation}
where $\bar{A}_{s}$ is the sample average of adjacency matrices in group $s$ as in \eqref{eq:A_bar},
$\bar{A}$ is the overall sample average of all the adjacency matrices,
    \[
        \bar{A} = \frac{1}{m} \sum_{s=1}^S \sum_{k = 1}^{m_s} A_s^{(k)} \enskip ,
    \]
$m$ is the total sample size,
    \[
        m = \sum_{s=1}^Sm_s \enskip ,
    \]
and $B$ is defined as in \eqref{eq:B}.

If $\theta^{(s)} = \frac{1}{\sqrt{15}} \trace\big((Z^{(s)})^3\big)$, then, under the null distribution and appropriate conditions on $m_s$, Theorem \ref{theorem theta hat unweight} gives
    \[
        \theta^{(s)}  \overset{d}{\to} \cN(0,1) \quad \text{as} \quad n \to \infty \enskip ,
    \]
and it follows that
    \[
        \big(\theta^{(s)}\big)^2 \overset{d}{\to} \chi^2(1) \quad \text{as} \quad n \to \infty \enskip .
    \]

Unfortunately, $\theta^{(1)}, \ldots, \theta^{(S)}$ are not independent, so the sum of their squares is not $\chi^2(S)$.
However, it is shown in \citet{ferrari2019note} that the sum of dependent $\chi^2$ random variables can be approximated with a gamma distribution.
Therefore, we have
    \begin{equation}
        \theta \equiv \sum_{s=1}^S \big(\theta^{(s)}\big)^2  \overset{d}{\to} \Gamma\Big(\frac{S}{u}, u\Big) \quad \text{as} \quad n \to \infty \enskip , \label{eq:anova}
    \end{equation}
where the scale parameter $u$ is given by
    \[
        u = 2\Big(1 + \frac{2 \sum_{q \neq r}^S \rho_{qr}}{S}\Big) \enskip ,
    \]
with $\rho_{qr}$ the pairwise correlation between $\big(\theta^{(q)}\big)^2$ and $\big(\theta^{(r)}\big)^2$.
    
As before, the true link probability matrices $P_s$ are unknown and need to be estimated.
We can estimate each $\hat P_s$ as in Section \ref{estimating link}, and then plug in these estimates to $Z^{(s)}$ in \eqref{anova}.
Furthermore, although the pairwise correlations $\rho_{qr}$ are not analytically tractable, they can easily be estimated using the Monte Carlo simulations in Algorithm \ref{alg:test}, which does not add to the computational complexity.

Moreover, with this setup, it is possible to follow the same development of Theorem \ref{theorem theta hat unweight} in proving convergence of the plug-in estimator $\hat \theta$ that uses the estimated link probability matrices and estimated pairwise correlations.
Similarly, \eqref{anova} can be extended to weighted networks as in Section \ref{sec-weighted}.

\section{Simulation studies}\label{sec-simulation}
    
In this section, we illustrate the performance of our proposed tests through extensive simulation study. 
For binary networks, we evaluate three plug-in estimators for the link probability matrices -- AVG, SBM, and MNBS -- and compare the results to the test proposed in \citet{GhoLux18}, which involves an estimated distance between two network distributions based on the Frobenius measure for binary networks that allows $n$ to go to infinity. 
We refer to these four tests as $\text{SPE-AVG}, \ \text{SPE-SBM}, \ \text{SPE-MNBS}$, and $\text{DFRO}$, respectively.
We do not include the test proposed in \citet{ginestet2017hypothesis} for comparison because their results are asymptotic in the sample size with a fixed number of nodes and the authors note that their test is expected to decrease power in larger dimensions, i.e. with more nodes.
    
In our simulation studies, we evaluate the test performance by estimating the power when the alternative is true, as well as the null rejection rate (rejection rate under the null).
We also vary the number of nodes, $n \in \{100, 200, \ldots, 1000\}$, and the sample sizes, $m_1 = m_2 = m \in \{10, 50\}$.
In each example, we set the significance level to $\alpha = 0.05$.
We follow the procedure described in Algorithm \ref{alg:test} with $Q = 1$ and report the empirical significance level as the average rejection rate on 5000 separate samples of networks from the underlying distributions.
Note that sampling new networks allows us to use $Q = 1$, but the results are similar if we use 5000 separate samples of networks with $Q > 1$.

With this design, we consider four types of random graph model for sampling binary and weighted networks, as well as one multiple-sample setting.
The results are detailed in the remainder of this section, but the conclusions are as follows.
Overall, it appears that SPE-MNBS is the most robust to different network structures and sample sizes.
If the networks are drawn from an SBM, then unsurprisingly SPE-SBM is suitable.
Throughout, we see that SPE-AVG shows significant improvement as the sample size increases.
Finally, all three plug-in estimates of the link probability matrices yield superior results for our test compared to DFRO.
    
\subsection{Stochastic block model (SBM)}

In the first example, we consider an SBM structure with a block matrix given as
\begin{align}P_{\text{SBM}} = 
    \left[
    \begin{matrix}
    	0.5 + \varepsilon_1 & 0.25 \\
    	0.25 & 0.5
    \end{matrix}
    \right] \enskip ,\label{eq:sbm}
\end{align}
where $\varepsilon_1$ depends on our hypothesis.
The membership of the $i$th node is
\[M(i) = I(1 \leq i \leq \lfloor n / 3 \rfloor) + 2 I(\lfloor n / 3 \rfloor + 1 \leq i \leq n) \enskip .\]

\begin{figure}[!htb] 
    \centering
    \includegraphics[width = 1\textwidth]{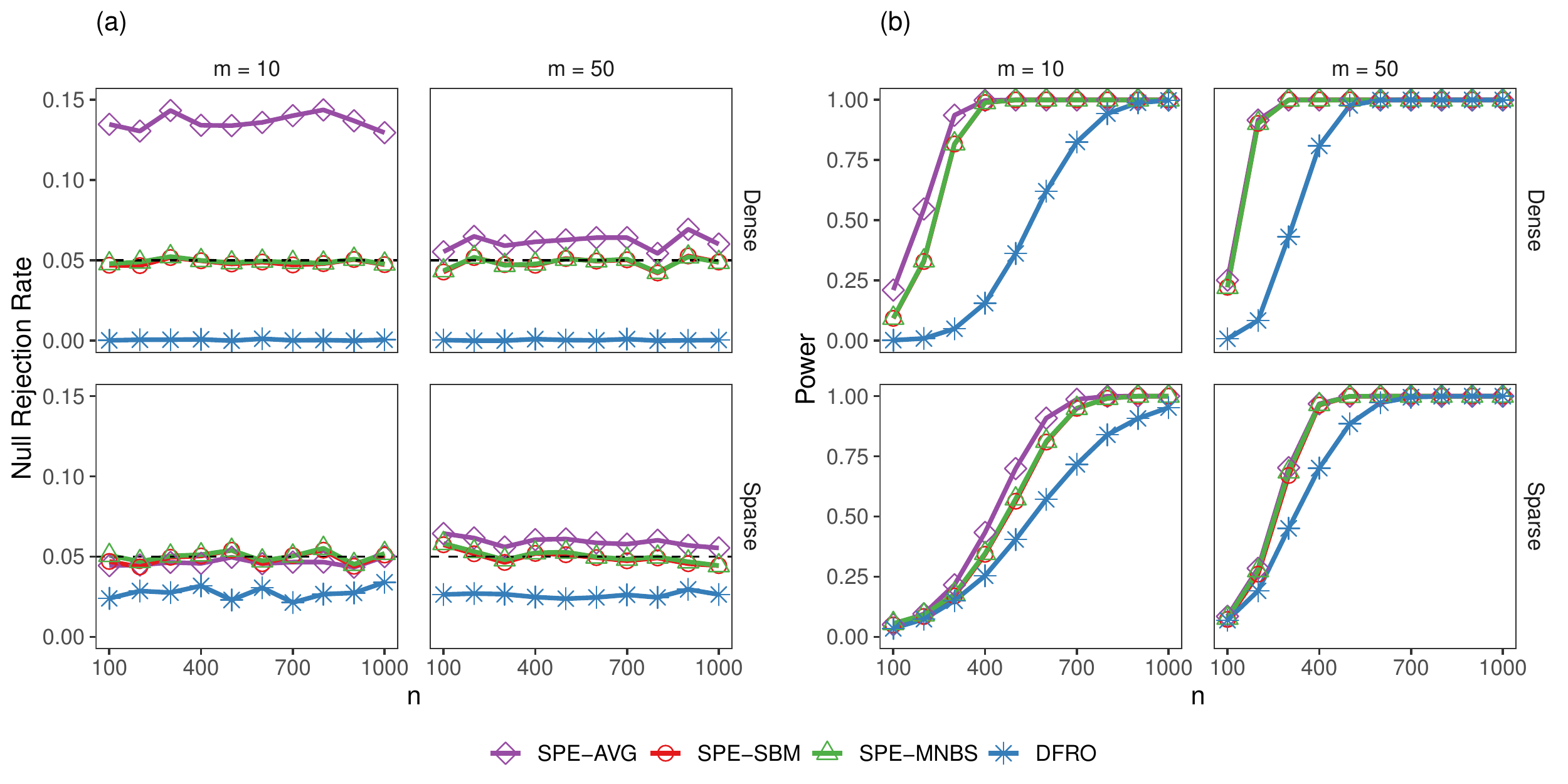} 
    \caption{Simulation results for testing networks with an SBM structure for different network orders and sample sizes.
    The first and the second rows are for dense and sparse networks, respectively.
    (a) Null rejection rate.
    (b) Power under the alternative.}
    \label{simulation 1}
\end{figure}

The first group of networks, $\{A_1^{(k)}\}_{k = 1}^{m_1}$, is generated from $P_{\text{SBM}}$ with $\varepsilon_1 = 0$.
In the null setting, the second group of networks, $\{A_2^{(k)}\}_{k = 1}^{m_2}$, is also generated from $P_{\text{SBM}}$ with $\varepsilon_1 = 0$, whereas $\varepsilon_1 = 1/(5 \log m)$ in the alternative setting.
The results are shown in the first row of Figure \ref{simulation 1}.
    
To investigate the performance of the tests for sparser networks, the same setting as above is considered, except now with $\varepsilon_1 = 2/(5 \log m)$ and with the link probability matrix $P_{\text{SBM}}$ scaled by a factor $\rho = 1 / 4$.
The corresponding results are shown in the second row of Figure \ref{simulation 1}. 
    	
We see from the first row of Figure \ref{simulation 1}, where the networks are dense, that SPE-SBM and SPE-MNBS are close to the nominal level $\alpha = 0.05$ under $H_0$ and both achieve good power under $H_1$.
We also observe that SPE-AVG is the most powerful under $H_1$, but its rejection rates are too high under $H_0$ when $m = 10$.
However, this issue is mitigated when we increase the sample size to $m = 50$ even though this makes $\varepsilon_1$ smaller, i.e. more similar underlying SBM structures.
For DFRO, it has zero rejection rate under $H_0$ and increases to unit power more slowly than our proposed tests.
Similar results hold in the sparser settings on the bottom row, though we note that SPE-AVG and DFRO perform more comparably to SPE-SBM and SPE-MNBS.
    	    	
\subsection{Graphon}

\begin{figure}[!htb] 
    \centering
    \includegraphics[width = 1\textwidth]{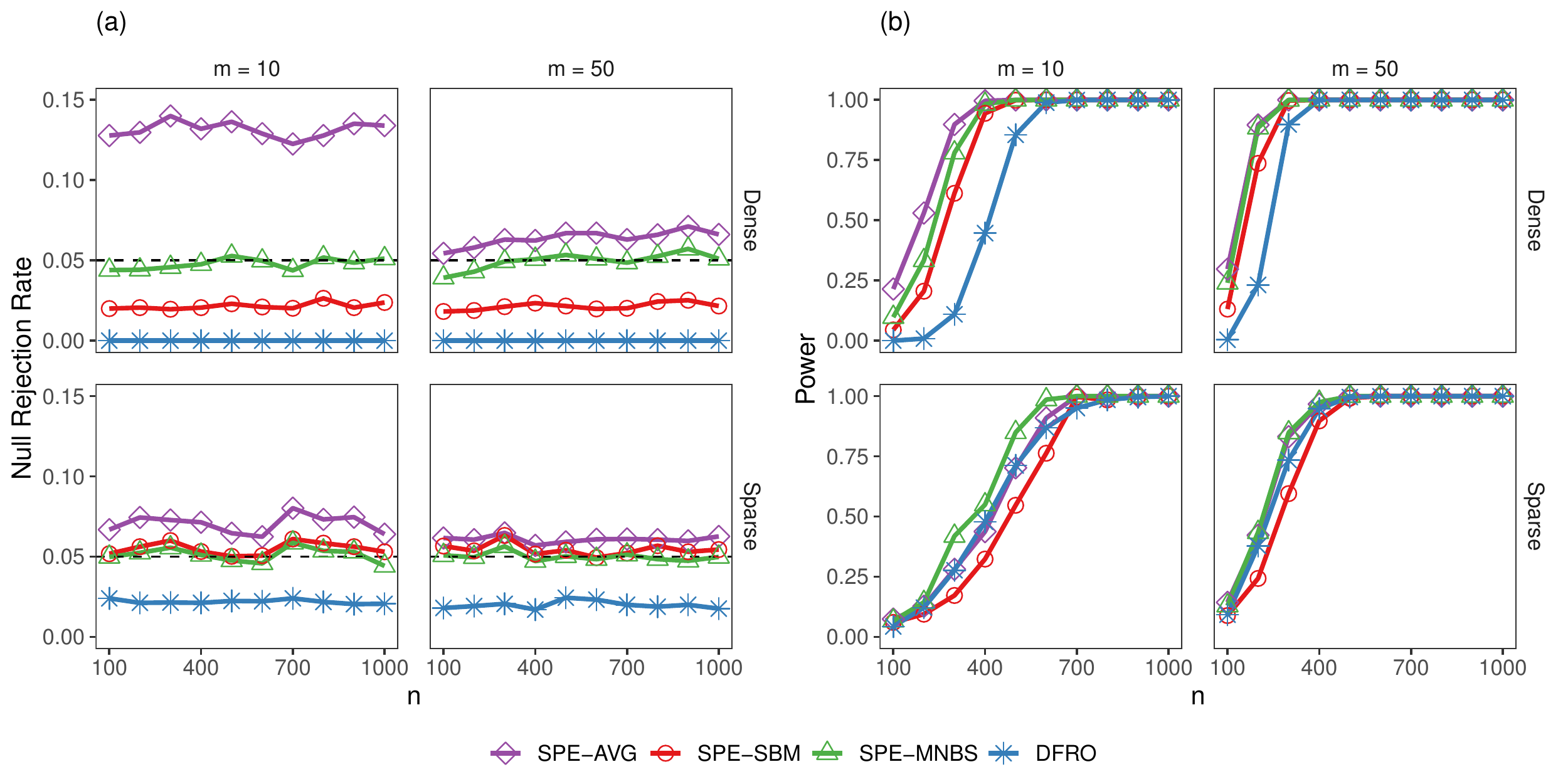} 
    \caption{Simulation results for testing networks with a graphon structure.
    The first and the second rows are for dense and sparse networks, respectively.
    (a) Null rejection rate.
    (b) Power under the alternative.}
    \label{simulation 2}
\end{figure}

In the second example, we consider a graphon structure from \citet{chan2014consistent} in which
\begin{align*}
    f_0(v_1,v_2) = (v_1^2 + v_2^2 + v_1^{1 / 2} + v_2^{1 / 2}) / 4 \enskip .
\end{align*}
Then $P_{u, i j} = f_0(\eta_i,\eta_j)$, with $\eta_i \overset{i.i.d}{\sim} U(0,1)$ for $i = 1, 2, \ldots, n$. 
We generate $\{A_1^{(k)}\}_{k = 1}^{m_1}$ from probability matrix $P_1$ according to $f_0$.
For the second group of networks, under the null hypothesis, we again sample from $f_0$ to generate $\{A_2^{(k)}\}_{k = 1}^{m_2}$.
Under the alternative hypothesis, we first randomly choose a subset $S \subset \{1, 2, \ldots, n\}$ with $\vert S \vert = \lfloor n / 10 \rfloor$, where $\lfloor \cdot \rfloor$ is the floor operator, then generate $\{A_2^{(k)}\}_{k = 1}^{m_2}$ from $P_2$ with $P_{2, ij} = P_{1, ij} - \varepsilon_2$, where
	\[\varepsilon_2 = \begin{cases}
	1 / (5 \log m) & \text{if } i, j \in S \\
	0 & \text{if } i, j \not\in S
	\end{cases} \enskip .\]
The results are presented in the first row of Figure \ref{simulation 2}.
As before, we set $\varepsilon_2 = 2 / (5 \log m)$ for $i, j \in S$ and scale the link probability matrix $P_1$ by a factor $\rho = 1 / 4$ to yield sparser networks.
The  results are shown in the second row of Figure \ref{simulation 2}. 
    	
We see from Figure \ref{simulation 2} that SPE-MNBS exhibits superior performance over the other tests in terms of both null rejection rate and power.
We also observe that SPE-SBM shows a lower rejection rate than the nominal level in the dense case, which suggests that SPE-SBM is more sensitive to network topologies that deviate from an SBM.
The behaviors of SPE-AVG and DFRO are similar to those in the first example, as we continue to see subpar performance, especially for small $m$.
    
\subsection{Correlated Erd\"os-R\'enyi model}

\begin{figure}[!htb]
 	\centering
 	\includegraphics[width=1\columnwidth]{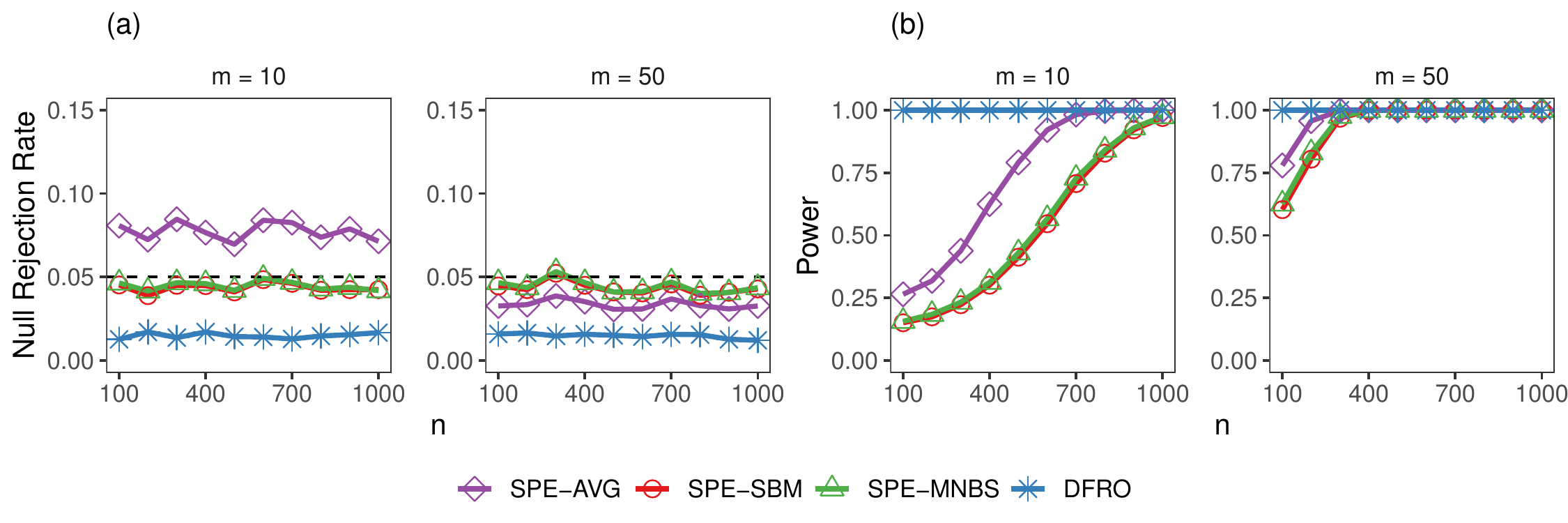} 
    \caption{Simulation results for testing networks with a correlated ER structure. 
    (a) Null rejection rate.
    (b) Power under the alternative.}
    \label{simulation 3}
 \end{figure}

In the third experiment, we study the robustness of the four tests to dependency.
For this, we consider the correlated Erd\H{o}s-R\'{e}nyi (ER) model from \citet{pedarsani2011privacy}.
We begin by sampling two independent Erd\H{o}s-R\'{e}nyi networks, $A_1 \sim ER(n, p_1)$ and $A_2 \sim ER(n, p_2)$.
We generate $\{A_1^{(k)}\}_{k = 1}^{m_1}$ with a parameter $\varepsilon_3$ as follows:
    \[A_{1, ij}^{(k)} \sim \begin{cases}
        \text{Bernoulli}(\varepsilon_3) &\text{ if } A_{1, i j} = 1\\
        0 &\text{ if } A_{1, i j} = 0
    \end{cases} \enskip .\]
This yields $m_1$ networks that are marginally $ER(n, p_1\varepsilon_3)$, but whose edge sets are correlated.
We similarly generate $\{A_2^{(k)}\}_{k = 1}^{m_2}$ conditional on $A_2$ with parameter $\varepsilon_4$.
We set $\varepsilon_3 = \varepsilon_4 = 0.8$ and $p_1 = 0.9$.
Under the null hypothesis, we set $p_2 = p_1 = 0.9$, whereas $p_2 = 0.83$ under the alternative hypothesis.
The results are shown in Figure \ref{simulation 3}.

We see that DFRO has consistently high power in the alternative setting for the entire range of $n$, which is only matched for our tests as $n$ increases, with SPE-AVG outperforming both SPE-SBM and SPE-MNBS.
However, the rejection rate under the null is below the nominal level for DFRO, whereas both SPE-SBM and SPE-MNBS are very close to $\alpha = 0.05$.
SPE-AVG has a higher rejection rate than expected when the sample size is $m = 10$, but this improves when $m = 50$.
Overall, it appears that SPE-SBM and SPE-MNBS are robust to the independence violation when $n$ is large.
    
\subsection{Beta weight distribution}

In our fourth experiment, we evaluate the tests on weighted networks by generating edge weights from Beta distributions.
Since SPE-MNBS and DFRO are for binary networks, we only consider SPE-AVG and SPE-SBM tests in this experiment. 

\begin{figure}[!htb] 
	\centering
	\includegraphics[width = 1\textwidth]{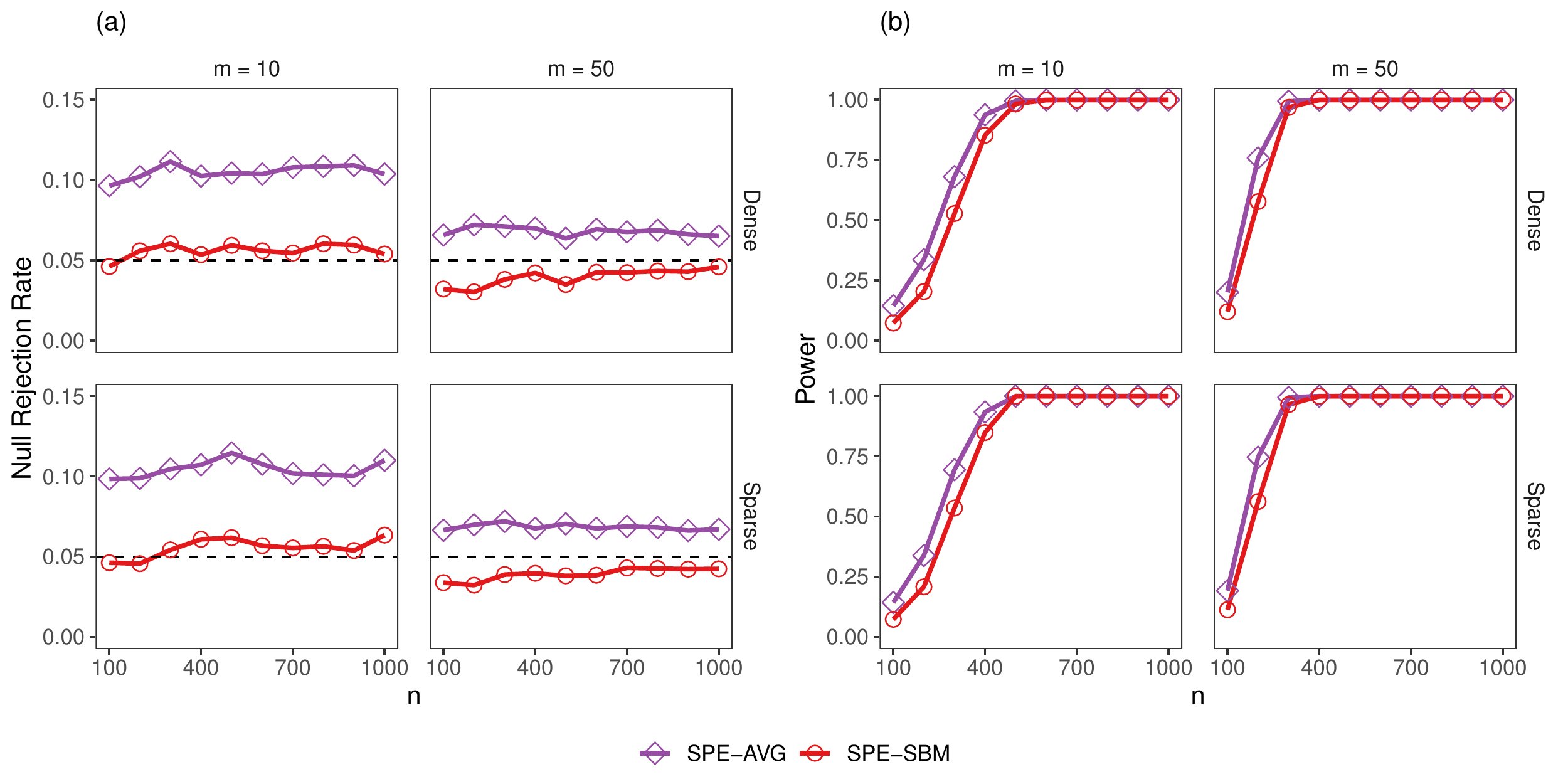} 
    \caption{Simulation results for testing weighted networks with a community structure and Beta-distributed edge weights.
    The first and the second rows are for dense and sparse networks, respectively.
    (a) Null rejection rate.
    (b) Power under the alternative.}
    \label{simulation 4}
\end{figure}

Each network is constructed with two communities, where the membership of the $i$th node is given by
    \[M(i) = I(1 \leq i \leq  n / 2 ) + 2 I( n / 2  + 1 \leq i \leq n) \enskip .\]
When nodes $i$ and $j$ are in the same community, $A_{1, i j}^{(k)} \sim \mathrm{Beta}(x_1,x_2)$ and $A_{2, i j}^{(k)} \sim \mathrm{Beta}(x_1 + \varepsilon_5,x_2 + \varepsilon_5)$.
Otherwise, $A_{1, i j}^{(k)} \sim \mathrm{Beta}(y_1,y_2)$ and $A_{2, i j}^{(k)} \sim \mathrm{Beta}(y_1 + \varepsilon_5,y_2 + \varepsilon_5)$, where $x_1 = 2, x_2 = 8, y_1 = 4, y_2 = 1$,  $1 \leq i < j \leq n$.
We set $\varepsilon_5 = 0$ and $\varepsilon_5 = 1 / (5 \log m)$ for the null and alternative hypotheses, respectively.
Again, we scale the weighted adjacency matrices by a factor $\rho = 1 / 4$ to obtain sparser networks.
The results are shown in Figure \ref{simulation 4}. 
    
Figure \ref{simulation 4} shows that SPE-SBM has good performance under both hypotheses even though the networks are weighted.
However, SPE-AVG suffers from a high type-one error especially when the sample size $m$ is small.
Both tests have slightly higher power in the dense setting, but the difference is small.

\subsection{Multiple sample testing}

\begin{figure}[!htb] 
 	\centering
 	\includegraphics[width = 1\textwidth]{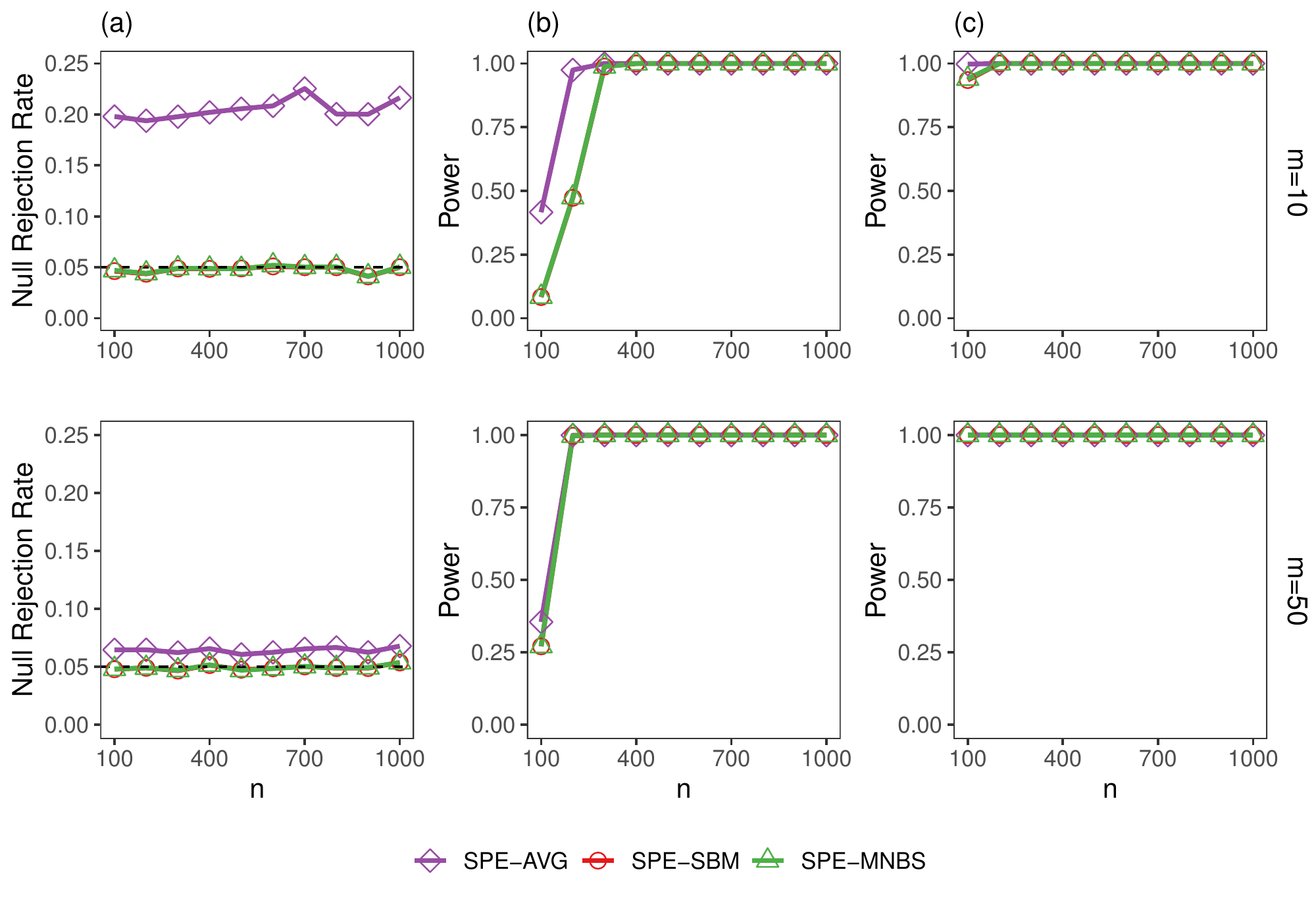} 
    \caption{Simulation results for testing multiple groups of binary networks with SBM community structure.
    The first row is $m = 10$ and the second row is $m = 50$.
    (a) Null rejection rate when all three groups are the same.
    (b) Power when two of the groups are the same and the third group is different.
    (c) Power when all three groups are different.}
    \label{simulation 5}
 \end{figure}

Finally, we evaluate our multiple-sample test from Section \ref{sec-anova} using the test statistic in \eqref{eq:anova}.
As in our first simulation, we consider the same SBM structure with block matrix given in \eqref{eq:sbm}. 
In the null case, we generate three groups, $\{A_s^{(k)}\}_{k = 1}^{m_s}$, $s = 1, 2, 3$, using $\varepsilon_1 = 0$. 
In the second case, we let the first two groups be sampled from $P_{\text{SBM}}$ with $\varepsilon_1 = 0$, while the third group uses $\varepsilon_1 = 1 / (5\log m)$.
In the third case, we sample the first group of networks from $P_{\text{SBM}}$ with $\varepsilon_1 = 0$, while the second and third groups are with $\varepsilon_1 = 1/(5 \log m)$ and $- 1/(5 \log m)$, respectively.
The results are given in Figure \ref{simulation 5}.

We see that SPE-AVG has the highest power for small $n$ in both alternative scenarios.
However, SPE-AVG has a very inflated null rejection rate for $m = 10$, which is significantly improved, though still slightly inflated, for $m = 50$.
On the other hand, SPE-SBM and SPE-MNBS have null rejection rates that oscillate around the nominal rate of $\alpha = 0.05$, indicating that the gamma approximation for the test statistic, which is a sum of dependent $\chi^2$ statistics, is very accurate.
Under the alternative, these methods perform similarly: when two groups are the same and one is different, the power quickly converges to one, and when all three groups are different, the power almost begins at one.

\section{Real data examples}\label{sec:realdata}

In this section, we apply our tests on three real datasets representing three different settings of interest to the biological research community.
The first two are networks constructed from fMRI data that represent two distinct streams of fMRI usage, the former being a case/control study and the latter being task-based.
The third dataset is derived from microbial measurements, an area in which network-based representations have recently emerged as a popular technique for studying the bacteria present within a microbiome \citep{layeghifard2017disentangling}.

In all three cases, the networks are weighted.
Therefore, we present results from our test for weighted networks in Section \ref{sec-weighted}.
To understand the performance of our tests for binary networks from Section \ref{sec-unweighted}, we also present the results as a function of thresholding the weights to binarize the networks (as is often done in practice).

\subsection{Data}

Our first dataset comes from the Center for Cognitive Brain Imaging at Carnegie Mellon University.
As part of the StarPlus experiment, fMRI data were collected for 6 individuals over 40 trials \citep{mitchell2004learning}.
During each trial of the experiment, an individual was shown an initial stimulus (either sentence or picture) for 4 seconds, then a blank screen for 4 seconds, and finally a second stimulus for up to 4 seconds. 
The trial ended when the individual answered whether the sentence correctly described the picture. 
After a rest period of 15 seconds, the trial was repeated with a new term. 
Images were recorded every 0.5 seconds and thus there are a total of 54 images for each trial which lasted for 27 seconds. 
In half of the 40 trials, the picture was presented first, and in the remaining trials, the sentence was presented first.
We refer to these two datasets as PS and SP, respectively.
The data are publicly available at \url{http://www.cs.cmu.edu/afs/cs.cmu.edu/project/theo-81/www/}.

Brain networks are constructed using the fMRI measurements from this study, where nodes correspond to the 24 regions of interest (ROIs) given by \citet{hutchinson2009modeling} and edges represent connectivity between the ROIs.
Specifically, for each time point $T = 1, \ldots, 54$, we combine the samples from the 6 individuals and trials in the PS or SP datasets, and estimate the correlation matrix by a shrinkage estimator proposed in \citet{schafer2005shrinkage}.
As a result, we obtain 54 networks from the PS dataset and 54 networks from the SP dataset, which we call NetPS and NetSP, respectively.
Finally, the fMRI time-series when the individuals view a sentence consist of the first 16 networks in NetSP and the 17th-32nd networks in NetPS.
The fMRI time-series when the individuals view a picture consist of the 17th-32nd networks in NetSP and the first 16 networks of NetPS. 
These final two groups are referred to as NetS and NetP, respectively.
     
The StarPlus networks derive from fMRI data in a task-based setting.  Another common use of fMRI data is case/control studies. One such example is the COBRE dataset from \citet{aine2017multimodal}, which consists of 124 brain networks for 54 schizophrenics and 70 controls.
These brain networks are also constructed from fMRI data, where nodes correspond to 264 ROIs defined in \citet{power2011functional} and edges represent the Pearson correlation coefficient between the time series of fMRI activity in the corresponding ROIs.
We use the networks constructed in \citet{relion2019network}, who provide further detail on the pre-processing and registration steps used to construct the networks.

The final dataset is from \citet{digiulio2015temporal}.
The authors tracked the microbiomes (MBs) of 37 women over the course of their pregnancies, 11 of which resulted in preterm delivery.
This study resulted in a time series of operational taxonomic unit (OTU) tables, which are counts of taxa present in each sample.
MB networks can be constructed from such data, where nodes correspond to taxa and edges represent the co-occurrence of taxa in the corresponding OTUs.
We use the networks constructed in \citet{josephs2020bayesian} and note that network analysis is an emerging tool in the MB literature.

\subsection{Results for weighted tests}
\label{subsec: weighted results} 
  
We begin by applying our tests for weighted networks from Section \ref{sec-weighted} to all three datasets.
We test if the groups defined by their respective labels -- picture/sentence, schizophrenic/control, preterm/term -- are different.
To do so, we specify a null hypothesis which says the underlying random distributions are equal against the alternative that says they are different.
We refer to this as the ``alternative setting” because the two samples are different with respect to their group label.


For $\alpha = 0.05$ and $Q = 1000$, we find that for the StarPlus networks, SPE-AVG and SPE-SBM correctly reject the null with reject rates of 1 and 0.71, separately, which are consistent with previous research on distinguishing the cognitive states of looking at a picture and a sentence \citep{mitchell2004learning, wang2004training, mitchell2003classifying}.
For both the COBRE and MB datasets, we find a rejection rate of one for both SPE-AVG and SPE-SBM.

Next, we perform an in silico experiment with the real data by subsampling within one of the classes. We refer to this as the ``null setting.”
The rationale for this setup is that we do not actually know if the different groups are generated by different underlying distributions, e.g one for schizophrenic and another for non-schizophrenic.
Therefore, we want to check if the null rejection rate is close to the nominal level in an experiment where all of the networks are 
from the same group.

To do so, we test the entire 
NetP, non-schizophrenic, and term delivery groups against a subsample (with half of the original sample size) of the same group for the StarPlus, COBRE, and MB datasets, respectively. 

After $1000$ random subsamples of the networks and $Q = 1$ for each subsample, for the StarPlus networks, SPE-SBM fails to reject the null hypothesis with reject rates of 0.003, which is expected since the samples are drawn from the same population.
However, SPE-AVG rejects the null with an inflated rate of 0.096. 
For the COBRE networks, we obtain null rejection rates of 0.78 and 0.68 for SPE-AVG and SPE-SBM, respectively.
The null rejection rates are improved for the MB networks with 0.63 and 0.46 for SPE-AVG and SPE-SBM, respectively, but still very inflated compared to the nominal $\alpha = 0.05$.

We speculate that this is happening because even within one class, there is a lot of variation.
That is, a subsample of brain networks with schizophrenia may look very different from another sample of brain networks with schizophrenia because we are not controlling for potential factors such as age and sex.
We refer to this issue as having too much heterogeneity within a class. This heterogeneity could lead to inflated null rejection rates because the underlying distributions of the two samples are different, but the difference is not the one we are trying to isolate.

\subsection{Results for binary tests}

As the results for the weighted tests showed inflated rejection rates in our simulated null setting, there is reason to believe the networks are too heterogeneous within each class.
Furthermore, many of the weights could represent spurious correlations.
Therefore, this is a setting in which binarizing the weights could improve the signal-to-noise ratio.
This idea is related to a common problem in the neuroscience literature related to the issue of sensitivity to thresholding edges \citep{ginestet2014statistical, garrison2015stability}.

\begin{figure}[!htb] 
 	\centering
 	\includegraphics[width = 1\textwidth]{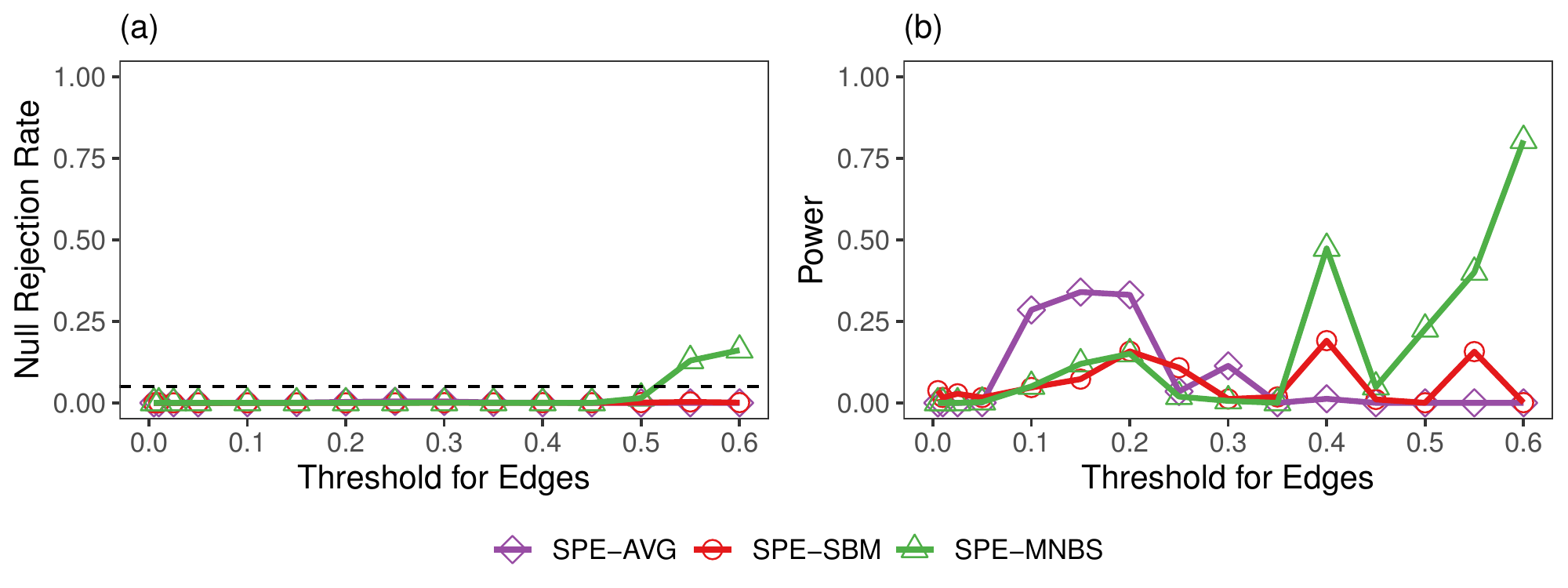}
 	\caption{Figures (a) and (b) show the null rejection rate and power, respectively, for different thresholds for binarizing the StarPlus networks.}
	\label{starplus}
 \end{figure}

\begin{figure}[!htb] 
	\centering
	\includegraphics[width = 1\textwidth]{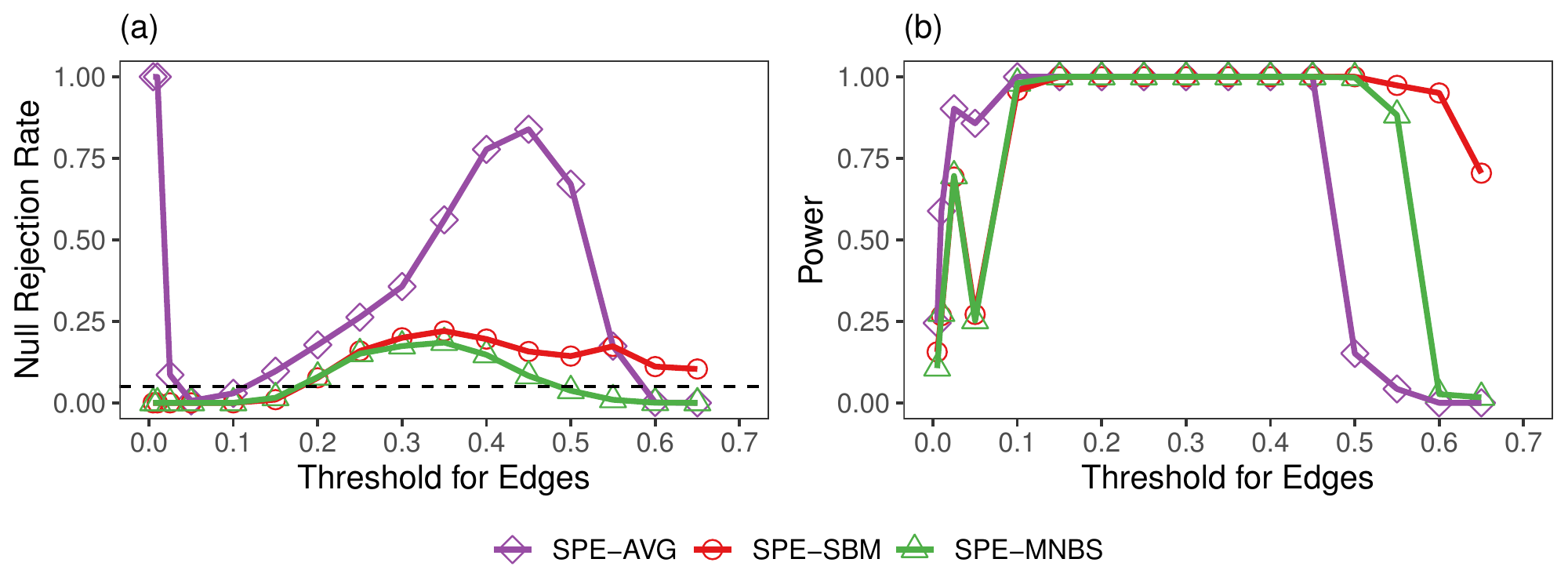} 
	\caption{Figures (a) and (b) show the null rejection rate and power, respectively, for different thresholds for binarizing the COBRE networks.}
	\label{COBRE}
\end{figure}

\begin{figure}[!htb] 
	\centering
	\includegraphics[width = 1\textwidth]{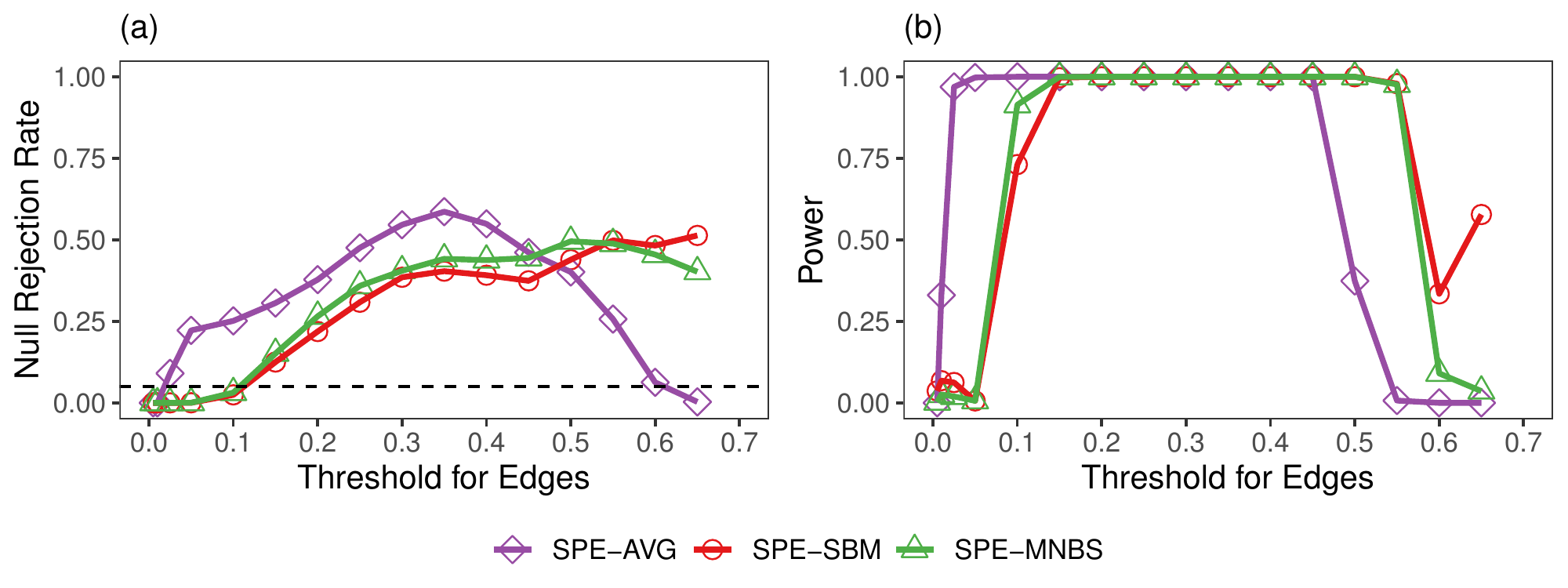} 
	\caption{Figures (a) and (b) show the null rejection rate and power, respectively, for different thresholds for binarizing the MB networks.}
	\label{MB}
\end{figure}

To evaluate this, we apply the binary tests from Section \ref{sec-unweighted} by binarizing the weights, which are all correlation values in $[-1, 1]$, based on thresholding their magnitude.
This threshold relates directly to the density of the networks.
With the same procedures in Section \ref{subsec: weighted results}, the results are given in Figures \ref{starplus}, \ref{COBRE}, and \ref{MB}.
The dashed lines for the null rejection rate in these figures all indicate the nominal level of 0.05.

The plots illustrate the tradeoff between the false positive rate in our null setting and the true positive rate in our alternative setting, which both are functions of the threshold.
As the threshold for an edge increases, the network becomes more sparse, resulting in a higher rejection rate in our null setting.
For thresholds above 0.6, some of the networks became too sparse, even resulting in some null graphs.
On the other hand, for a low threshold, there is less power to detect a difference in our alternative setting.
Such curves as a function of threshold could provide practitioners a way to understand the signal-to-noise ratio of their edge weights.

For the COBRE and MB networks, we have high power for a wide range of threshold values, which is consistent with our findings using the weighted networks directly.
However, we also see a low rejection rate in our null setting, especially for a threshold of 0.1, which seems to provide the best tradeoff.
This suggests that the signal-to-noise ratio in the weights is too low, which can be mitigated through thresholding.
For the StarPlus networks, a threshold between 0.1 and 0.2 seems to provide the best balance between signal and noise for SPE-AVG, whereas 0.55 is better for SPE-SBM and SPE-MNBS.

\section{Discussion}\label{discussion}

In this work, we proposed a new spectral-based statistic for hypothesis testing of populations of networks, which applies to both binary and weighted networks under a very general framework.
The test statistics are simple, computationally friendly, and theoretically supported by our derivation of both the limiting null distribution and asymptotic power guarantees.
We have demonstrated our method through extensive simulation study as well as real data analysis.
Future work will focus on exploring spectral-based methods for studying inference problems for networks with additional constraints or structures such as directed networks.

\printbibliography

\appendix

\section{Background on spectral results}\label{appendix: background}

Here, we review some results on spectral properties of inhomogeneous networks.
We begin with the generalized Wigner matrix. 

\begin{definition}[Generalized Wigner matrix \citep{bai2010spectral}]
	A generalized Wigner matrix is a Hermitian random matrix $W$ whose entries on or above the diagonal are independent.
\end{definition}

Let $W$ be an $n \times n$ generalized Wigner matrix with eigenvalues $\lambda_i$ for $i = 1,\cdots,n$.
The empirical spectral distribution (ESD) of $W$ is defined as
    \[
	    F_n(x)  = \frac{1}{n} \#\{\lambda_i \leq x, 1 \leq i \leq n\} \enskip ,
    \]
where $\#$ denotes counting measure. It was proved  in \citet{bai2010spectral} that under some assumptions, the ESD of a normalized $W$ approximates the semicircle law $F$:
    \[
	    F(x) = \frac{1}{\sqrt{2 \pi}} \sqrt{4 - x^2}, \ - 2 \leq x \leq 2 \enskip .
    \]
Specifically, the result is stated in the following lemma.
	
\begin{lemma}[Theorem 2.9 in \citet{bai2010spectral}]\label{Lemma 2.9}
	Suppose that $W = \frac{1}{\sqrt{n}} X$ is an $n \times n$ Wigner matrix and the entries above or on the diagonal of $X$ are independent but may be dependent on $n$ and may not be necessarily identically distributed.
	Assume that all the entries of $X$ are of mean zero and variance 1, and satisfy the condition that, for any constant $\eta > 0$, 
	\begin{equation}\label{condition of lemma 2.9}
		\lim_{n \to \infty} \frac{1}{n^2} \sum_{i j} \mathrm{E}(|X_{i j}|^2) I(|X_{i j}| \geq \eta \sqrt{n}) = 0 \enskip .
	\end{equation}
	Then the ESD of $W$ converges to the semicircular law almost surely.
\end{lemma}

Next, we give a useful theorem focusing on $F_n$, the ESD of a normalized $W$ in the following.
	
Let $\cU$ be an open set of the real line that contains the interval $[- 2, 2]$, which is the support of the semicircular law $F$.
Define $\cF$, the set of functions $f: \cU \to \R$, where $\R$ is the set of real numbers.
Consider the empirical process $G_n = \{G_n(f)\}$ given by:
	\begin{equation}\label{process G_n}
		G_n(f) = n \int_{-\infty}^{\infty} f(x) [F_n - F] d x, \quad f \in \cF \enskip .
	\end{equation}
It is shown in  \citet{bai2010spectral} that $G_n$ converges to a Gaussian process under the following moment conditions:
\begin{enumerate}
	\item[(1)] For all $i$, $\mathrm{E} (|X_{ii}|^2) = \sigma^2 > 0$, for all $i < j$, $\mathrm{E} (|X_{i j}|^2) = 1$.
	\item[(2)] (Homogeneity of fourth moments) $M = \mathrm{E} (|X_{i j}|^4)$ for $i \neq j$.
	\item[(3)] (Uniform tails) For any $\eta > 0$, as $n \to \infty$, 
	    \[
		    \lim_{n \to \infty} \frac{1}{\eta^4 n^2} \sum_{i j} \mathrm{E}\big(|X_{i j}|^4 I(|X_{i j}| \geq \eta \sqrt{n})\big) = 0 \enskip .
	    \]
\end{enumerate}
    
For any $f \in \cF$ and any integer $l \geq 0$, define
    \[
    	\tau_l(f) = \frac{1}{2 \pi} \int_{-\pi}^{\pi} f(2 \cos(\theta)) e^{i l \theta} d \theta \enskip .
    \]
Setting $\beta = \mathrm{E}((X_{1 2} - 1)^2) - 2$, there is the following theorem on the convergence of $G_n$.

\begin{lemma}[Theorem 9.2 in \citet{bai2010spectral}]\label{Lemma 9.2}
    Under conditions $(1)$--$(3)$, the spectral empirical process $G_n =( G_n(f))$ indexed by the set of analytic functions $\cF$ converges weakly in finite dimension to a Gaussian process $G = \{G(f), f \in \cF\}$ with mean function $\mathrm{E}{(G(f))}$ given by 
    	\[\frac{1}{4} (f(2) + f(-2)) - \frac{1}{2} \tau_0(f) + (\sigma^2 - 2) \tau_2(f) + \beta \tau_4(f) \enskip ,\]
    and the covariance function $c(f,g) = \mathrm{E}\big((G(f) - \mathrm{E}(G(f)) (G(g) - \mathrm{E}(G(g)) \big)$ given by 
    \[\frac{1}{4 \pi^2} \int_{-2}^2 \int_{-2}^2 f'(t) g'(t) V(t,s) d t d s \enskip ,\]
    where 
    \[V(t,s) = (\sigma^2 - 2 + \frac{1}{2} \beta t s) \sqrt{(4 - t^2) (4 - s^2)} + 2 \log \left(\frac{4 - t s + \sqrt{(4 - t^2) (4 - s^2)}}{4 - t s - \sqrt{(4 - t^2) (4 - s^2)}}\right) \enskip .\]
\end{lemma}

\section{Proofs of theorems from Section \ref{sec-test}}

\subsection{Proof of Theorem \ref{theorem theta unweight}}\label{proof: theta unweight}
    One can see that under the null hypothesis of $P_1 = P_2$, $Z$ is a Wigner matrix satisfying $\mathrm{E}(Z_{i j}) = 0$ and $\text{Var}(Z_{i j}) = 1 / n$. 
    The remainder of the proof proceeds in three steps.
    
    First, we show that the ESD of $Z$ converges to the semicircular law $F$ almost surely.
    Let $X = \sqrt{n} Z$.
    From Lemma \ref{Lemma 2.9}, it is sufficient to prove that condition \eqref{condition of lemma 2.9} is satisfied. For any constant $\eta > 0$, we have
   	    \begin{align*}
   	        & \quad \frac{1}{n^2} \sum_{i j} \mathrm{E}(|X_{i j}|^2) I(|X_{i j}| \geq \eta \sqrt{n}) \\
   	        & = \frac{1}{n^2} \sum_{i j} I(|X_{i j}| \geq \eta \sqrt{n}) \\
   	        & \leq \max_{i j} \{ I(|X_{i j}| \geq \eta \sqrt{n}) \} \\
            & = \max_{i \neq j} \{ I \left(|\bar{A}_{1, i j} - \bar{A}_{2, i j}| \geq \eta T \right), I(1 \geq \eta \sqrt{n})\} \\
   	        & \quad \to 0, \quad \text{as} \ n \to \infty \enskip ,
   	    \end{align*}
    where $T=\sqrt{n \left(\frac{1}{m_1} P_{1, i j} (1 - P_{1, i j}) + \frac{1}{m_2} P_{2, i j} (1 - P_{2, i j})\right)}$, 
    the last equality is due to $X_{i i} = 1$, and the last convergence comes from the boundedness of $|\bar{A}_{1, i j} - \bar{A}_{2, i j}|$, whereas $n/m_u \to \infty$  for $u = 1,2$ and $\eta \sqrt{n} \to \infty$.
   	
    Next, we verify that $X$ satisfies conditions $(1)$--$(3)$ of Lemma \ref{Lemma 9.2}. For simplicity, we still denote the ESD of $Z$ as $F_n$. Then, the empirical process $G_n$ defined in \eqref{process G_n} converges to a Gaussian process from Lemma \ref{Lemma 9.2} and the asymptotic normality of $\theta$ given by \eqref{theta} follows directly.
   	
    Condition $(1)$ holds since $\mathrm{E}(|X_{i j}|^2) = n \mathrm{E}(|Z_{i j}|^2) = 1$ for all $i$ and $j$.
   	
    By the central limit theorem, if $i\neq j$, then $Z_{i j}$ converges to the same normal distribution $\cN(0,1/n)$ when $m_1$ and $m_2$ are large.
    Thus condition $(2)$ is satisfied under the large sample size assumption.
   	
    Finally, we verify condition $(3)$ by applying a H\"{o}lder inequality. For any $\eta > 0$,
   	\begin{align*}
   	    & \quad \frac{1}{\eta^4 n^2} \sum_{i j} \mathrm{E}\big(|X_{i j}|^4 I(|X_{i j}| \geq \eta \sqrt{n})\big) \\
   	    & \leq \frac{1}{\eta^4 n^2} \sum_{i j} \big(\mathrm{E}(|X_{i j}|^8)\big)^{1 / 2}  \big(\mathrm{E} \big(I(|X_{i j}| \geq \eta \sqrt{n})\big)\big)^{1 / 2} \\
   	    & \leq \frac{C_1}{\eta^4 n^2} \sum_{i j} \big(P \big(|X_{i j}| \geq \eta \sqrt{n}\big)\big)^{1 / 2} \\
   	    & \leq \frac{C_1}{\eta^4} \max_{i \neq j} \left\{ \big(P \big(|X_{i j}| \geq \eta \sqrt{n}\big)\big)^{1 / 2},  \big(P \big(1 \geq \eta \sqrt{n}\big)\big)^{1 / 2} \right\} \\
        & = \frac{C_1}{\eta^4} \max_{i \neq j} \left\{ \left(P \left(|\bar{A}_{1, i j} - \bar{A}_{2, i j}| \geq \eta T \right)\right)^{1 / 2}, \big(P \big(1 \geq \eta \sqrt{n}\big)\big)^{1 / 2} \right\} \\
   	    & \quad \to 0, \quad \text{as} \ n \to \infty \enskip ,
   	\end{align*}
    where 
    $C_1$ is the upper bound of $\big(\mathrm{E}(|X_{i j}|^8)\big)^{1 / 2}$ and is a positive constant. The last convergence also follows from the boundedness of  $|\bar{A}_{1, i j} - \bar{A}_{2, i j}|$ and the infinity trend of $T$ and $\eta \sqrt{n}$ as $n$ tends to infinity.  	
   	
    Taking $f(x) = x^3$ and combining Lemma \ref{Lemma 9.2} with \eqref{process G_n}, we have
   	    \begin{align}
   	        G_n(f) & = n \int_{-\infty}^{\infty} f(x) [F_n - F] d x \nonumber\\
   	        & = n \int_{-\infty}^{\infty} f(x) F_n d x - n \int_{-\infty}^{\infty} f(x) F d x \nonumber\\
   	        & = \sum_{i = 1}^n f(\lambda_i (Z))\label{G_n} \enskip ,
   	    \end{align} 
   	    where $\lambda_i (Z)$ is the $i$th eigenvalue of $Z$. 
    The third equality holds because the first term is the expectation of $f(x)$ with respect to the counting measure $F_n$ and the second term is the integral of an odd function $f(x)F$.
   	
    By diagonalizing $Z = U^T \Lambda U$, where $U$ is some orthogonal matrix and $\Lambda$ is a diagonal matrix whose diagonal entries are the eigenvalues of $Z$. From \eqref{G_n}, we obtain
   	    \begin{align*}
   	        G_n(f)  &= \sum_{i = 1}^n f(\Lambda_{i i}) 
   	        = \trace(f(\Lambda)) 
   	        = \trace(U^T f(\Lambda) U) \\
   	        & = \trace( f(U^T \Lambda U)) 
   	        = \trace( f(Z)) 
   	        = \trace(Z^3) \enskip .
   	\end{align*}

    $G(x^3)$ is the same as in \citet{dong2020spectral}, and hence its mean and variance are known to be $\mathrm{E}(G(x^3)) = 0$ and $\text{Var}(G(x^3)) = 15$, respectively, which completes the proof.

\subsection{Proof of Theorem \ref{theorem theta hat unweight}}\label{proof: theta hat unweight}
    First, for any $i \neq j$, we have
        \begin{equation}\label{Z_hat and Z unweight}
    	    \hat{Z}_{i j} = \frac{\sqrt{\frac{1}{m_1} P_{1, i j} (1 - P_{1, i j}) + \frac{1}{m_2} P_{2, i j} (1 - P_{2, i j})}}{\sqrt{\frac{1}{m_1} \hat P_{1, i j} (1 - \hat P_{1, i j}) + \frac{1}{m_2} \hat P_{2, i j} (1 - \hat P_{2, i j})}} Z_{i j}  \enskip .
    	\end{equation}
    Recall that under the null hypothesis, $P_1 = P_2$.
    By Taylor expanding the numerator in \eqref{Z_hat and Z unweight}, we obtain
        \begin{align*} 
    	    & \quad \sqrt{\frac{1}{m_1} P_{1, i j} (1 - P_{1, i j}) + \frac{1}{m_2} P_{2, i j} (1 - P_{2, i j})} \\
    	    & = \sqrt{\frac{m_1 + m_2}{m_1 m_2}} \sqrt{ P_{1, i j} (1 - P_{1, i j})} \\
    	    & = \sqrt{\frac{m_1 + m_2}{m_1 m_2}} \Big( \sqrt{\hat{P}_{1, i j} \big(1 - \hat{P}_{1, i j}\big)} + O\big(P_{1, i j} - \hat{P}_{1, i j}\big)\Big) \\ 
    	    & \leq  \sqrt{\frac{m_1 + m_2}{m_1 m_2}} \Big( \sqrt{\hat{P}_{1, i j} \big(1 - \hat{P}_{1, i j}\big)} +  o_p(1) \Big) \\
    	    & = \sqrt{\frac{1}{m_1} \hat{P}_{1, i j} \big(1 - \hat{P}_{1, i j}\big) + \frac{1}{m_2} \hat{P}_{1, i j} \big(1 - \hat{P}_{1, i j}\big)} + \sqrt{\frac{m_1 + m_2}{m_1 m_2}}   o_p(1) \enskip ,
    	\end{align*}
    where the third equality is obtained under the assumption that $\max_{i j}|\hat{P}_{u, i j} - P_{u, i j}| =  o_p(1)$.
    
    Without loss of generality, assume $\hat{P}_{1, i j} \big(1 - \hat{P}_{1, i j}\big) \leq \hat{P}_{2, i j} \big(1 - \hat{P}_{2, i j}\big)$.
    Then
        \begin{align}\label{approx1 unweight}
    	    & \quad \sqrt{ \frac{1}{m_1} P_{1, i j} (1 - P_{1, i j}) + \frac{1}{m_2} P_{2, i j} (1 - P_{2, i j}) }  \\
    	    & \leq \sqrt{ \frac{1}{m_1} \hat{P}_{1, i j} \big(1 - \hat{P}_{1, i j}\big) + \frac{1}{m_2} \hat{P}_{2, i j} \big(1 - \hat{P}_{2, i j}\big) } + \sqrt{\frac{m_1 + m_2}{m_1 m_2}}  o_p(1) \enskip . \nonumber
    	\end{align}	
    Similarly, we have
    	\begin{equation}\label{approx2 unweight}
    	    \begin{split}		
    	        & \quad \sqrt{  \frac{1}{m_1} \hat{P}_{1, i j} \big(1 - \hat{P}_{1, i j}\big) + \frac{1}{m_2} \hat{P}_{2, i j} \big(1 - \hat{P}_{2, i j}\big)  }  \\
    	        & \leq \sqrt{  \frac{1}{m_1} P_{1, i j} (1 - P_{1, i j}) + \frac{1}{m_2} P_{2, i j} (1 - P_{2, i j})  } + \sqrt{\frac{m_1 + m_2}{m_1 m_2}}  o_p(1) \enskip .
    	    \end{split}
    	\end{equation}
    From \eqref{approx1 unweight} and \eqref{approx2 unweight}, we have
    	\begin{equation}\label{approx 3 unweight}
    	    \begin{split}
    	        & \quad \sqrt{  \frac{1}{m_1} P_{1, i j} (1 - P_{1, i j}) + \frac{1}{m_2} P_{2, i j} (1 - P_{2, i j})  }  \\
    	        & = \sqrt{  \frac{1}{m_1} \hat{P}_{1, i j} \big(1 - \hat{P}_{1, i j}\big) + \frac{1}{m_2} \hat{P}_{2, i j} \big(1 - \hat{P}_{2, i j}\big)  } + \sqrt{ \frac{m_1 + m_2}{m_1 m_2} }  o_p(1) \enskip .
    	       \end{split}
    	\end{equation}
    Then, combining \eqref{approx 3 unweight} with \eqref{Z_hat and Z unweight}, and noticing that when $i = j$, $\hat{Z}_{i j} = Z_{i j} = B_{i j}$, we can write
    	\begin{equation}\label{Z hat Z unweight}
    	    \hat{Z} - B =   (J + H) \circ (Z - B) \enskip ,
    	\end{equation}
    where $\circ$ denotes the Hadamard product, $J$ is the $n \times n$ all-ones matrix, and $H$ is an $n \times n$ matrix with entries $H_{i j} = \max_{u = 1, 2} O(\hat P_{u, i j} - P_{i j}) =  o_p(1)$.
    	
    Moreover, $H \circ B$ is a zero matrix, thus an immediate consequence of \eqref{Z hat Z unweight} is 
    	\[
    	    \hat{Z} = Z \circ (J + H) \enskip .
    	\]

    	We further have 
    	\begin{equation*}
    	\trace(\hat{Z})^3 - \trace(Z^3) = 3 \trace\big( Z^2 (Z \circ H)\big) + 3 \trace \big(Z (Z \circ H)^2\big) + \trace \big((Z \circ H)^3 \big) \enskip .
    	\end{equation*}
    	
    	Next, we prove that $\trace\big( Z^2 (Z \circ H)\big) = \sum_{i, k, l} Z_{i k} Z_{k l} Z_{l i} H_{l i} = o_p(1)$. The summation can be divided into two parts in terms of $(i,k,l)$. In particular, let $\{1, \cdots, n\}^3 = S_{1} \cup S_{2}, \ S_{1} \cap S_{2} = \emptyset$. In set $S_{1}$, $Z_{i k} Z_{k l} Z_{l i} \geq 0$, and in set $S_{2}$, $Z_{i k} Z_{k l} Z_{l i} < 0$.
    		\begin{align}
    	    	\trace\big( Z^2 (Z \circ H)\big) & = \sum_{i, k, l \in S_1} Z_{i k} Z_{k l} Z_{l i} H_{l i} + \sum_{i, k, l \in S_2} Z_{i k} Z_{k l} Z_{l i} H_{l i} \nonumber \\
    	    	& \leq \sum_{i, k, l \in S_1} Z_{i k} Z_{k l} Z_{l i} \max_{l i} \{H_{l i}\} + \sum_{i, k, l \in S_2} Z_{i k} Z_{k l} Z_{l i} \min_{l i} \{ H_{l i}\} \nonumber \\
    	    	& = o_p(1) \enskip . \label{ineq1}
    	    \end{align}
    	The last equality comes from as following. From Theorem \ref{theorem theta unweight}, $\trace (Z^3) = O_p(1)$, when $S_1 = \emptyset$ or $S_2 = \emptyset$, which means $Z_{i j} < 0$ or $Z_{i j} \geq 0$ for any $i$ and $j$, inequality \eqref{ineq1} holds straightforwardly. When $S_1 \neq \emptyset$ and $S_2 \neq \emptyset$, $\sum_{i, k, l \in S_1} Z_{i k} Z_{k l} Z_{l i}$  can be viewed as a part of the total sum $\sum_{i, k, l } Z_{i k} Z_{k l} Z_{l i}$ when all $Z_{i j} \geq 0$, so $\sum_{i, k, l \in S_1} Z_{i k} Z_{k l} Z_{l i} = O_p(1)$. Similarly, $\sum_{i, k, l \in S_2} Z_{i k} Z_{k l} Z_{l i} = O_p(1)$. Thus, inequality \eqref{ineq1} still holds.
    	
    	By the similar method, we have $\trace \big(Z (Z \circ H)^2\big) = o_p(1)$ and $\trace \big((Z \circ H)^3 \big) = o_p(1)$. As a result, 
    	\[
    		\trace(\hat{Z})^3 = \trace(Z^3) + o_p(1) \to \trace(Z^3) \quad \text{as} \quad n \to \infty \enskip, 
    	\]
    and our proof is complete.

\subsection{Proof of Theorem \ref{theorem power unweight}}\label{proof: power unweight}
	As in the proof of Theorem \ref{theorem theta hat unweight}, $\hat{Z} = Z \circ (J + H)$ and  
	    \begin{equation}\label{trace4}
	        \trace(\hat{Z}^3) = \trace(Z^3)  
	        + o_p(1) \enskip .
	    \end{equation}
	Let $\tilde{Z}$ be an $n \times n$ matrix with entries 
	    \[     	  	
	        \tilde{Z}_{i j} =
	            \begin{cases}
	                \frac{(\bar{A}_{1, i j} - \bar{A}_{2, i j}) - (P_{1, i j} - P_{2, i j})}{\sqrt{n \left( \frac{1}{m_1} P_{1,i j} (1 - P_{1,i j}) + \frac{1}{m_2} P_{2,i j} (1 - P_{2,i j})\right)}} & \text{ if } i \neq j \\
	                B_{i j} & \text{ if } i = j
	            \end{cases} \enskip .
	   \]
	Noting the definitions of $Z$ and $Z''$ in \eqref{Z} and \eqref{Z''}, respectively, it is obvious that $\tilde{Z} = Z - Z''$ and $\tilde{Z}$ is a Wigner matrix.
	Thus,
	    \begin{align}
	        \trace(\tilde{Z}^3 ) = \trace(Z^3 - (Z'')^3 + 3 Z (Z'')^2 - 3 Z^2 Z'' ) \enskip .
	    \end{align}
	
	First, similar with the proof of Theorem \ref{theorem theta}, we have
	    \begin{equation}\label{trace1}
	        \trace(\tilde{Z}^3)  / \sqrt{15} \overset{d}{\to}  \cN(0,1) \enskip .
	    \end{equation}  
	
	Next, we calculate $\trace(3 Z (Z'')^2 - 3 Z^2 Z'')$:
	    \begin{align} \label{equa1}
	        \trace( 3 Z (Z'')^2 - 3 Z^2 Z'') = 3 \sum_{i, k, l} Z_{i k} Z''_{k l} Z''_{l i} - 3 \sum_{i, k, l} Z_{i k} Z_{k l} Z''_{l i} \enskip .
	    \end{align}
	When all $i, \ k, \ l$ are all different, $\mathrm{E}(Z_{i k} Z''_{k l} Z''_{l i} - Z_{i k} Z_{k l} Z''_{l i}) = 0$ due to the independence of the elements and the fact that $\mathrm{E}(Z_{i j}) = Z''_{i j}$ for all $i$ and $j$.
	When at least two subscripts are the same, we note that $\mathrm{E}(Z_{i i}) = Z''_{i i} = 0$.
	Therefore, $\mathrm{E}(Z_{i k} Z_{k l} Z''_{l i} -  Z_{i k} Z''_{k l} Z''_{l i}) = 0$ still holds, and hence $\mathrm{E}(\trace(3 Z (Z'')^2 - 3 Z^2 Z'')) = 0$.
	
	Next, from \eqref{equa1}, $\text{Var}(\trace( 3 Z^2 Z'' - 3 Z (Z'')^2)) = \max_{u = 1, 2}O(m_u^2)$.
	Using Bernstein's inequality, we have
    	\begin{equation}\label{trace2}
	        \trace(3 Z (Z'')^2 - 3 Z^2 Z'') = \max_{u = 1, 2} O_p(m_u^2) \enskip .
    	\end{equation}
	
	Finally, we have
	    \begin{align}
	        |\trace((Z'')^3)| &= \left|\sum_{i, k, l} Z''_{i k} Z''_{k l} Z''_{l i}\right| \enskip . \label{sum3}
	    \end{align}
	The sum in \eqref{sum3} can be divided into two parts in terms of the sets $S_a$ and $S_b$ as
	    \begin{align*}
	        |\trace((Z'')^3)| &\geq \left|\sum_{(i, k, l) \in S_a} Z''_{i k} Z''_{k l} Z''_{l i}\right| - \left|\sum_{(i, k, l) \in S_b} Z''_{i k} Z''_{k l} Z''_{l i}\right| \\
	        & \geq a n^3 \min_{(i, k, l) \in S_a} |{Z''}_{i k}|^3 + b n^3 \min_{(i, k, l) \in S_b} |{Z''}_{i k}|^3 > 0, \enskip
	    \end{align*}
	    where with a slight abuse of notation, $\min \{Z_{ik}\}$ is taken over all the pairs of indices among $(i,k,l).$
	It is obvious to see $|\trace((Z'')^3)| \geq \max_{u = 1, 2} O((m_u n)^{3 / 2})$.
	
	Similarly, we can write
	    \begin{align*}
	        |\trace((Z'')^3)| \geq - b n^3 \max_{(i, k, l) \in S_b} |{Z''}_{i k}|^3 - a n^3 \max_{(i, k, l) \in S_a} |{Z''}_{i k}|^3 > 0 \enskip .
	    \end{align*}
	Again have $|\trace((Z'')^3)| \geq \max_{u = 1, 2} O((m_u n)^{3 / 2})$.
	Therefore, in either case, we see that
	    \begin{equation}\label{trace3}
	        |\trace((Z'')^3)| \geq \max_{u = 1, 2} O((m_u n)^{3 / 2}) \enskip .
	    \end{equation}
	
	Combining \eqref{trace1}, \eqref{trace2} and \eqref{trace3}, we are ready to prove the asymptotic power of our test.
	    \begin{align*}
	        & \quad P(|\theta| > \mu_{\alpha / 2}) \\
	        & = P\left(\frac{1}{\sqrt{15}}|\trace(Z^3)| > \mu_{\alpha / 2}\right) \\
	        & = P\left(\frac{1}{\sqrt{15}}|\trace\big(\tilde{Z}^3 + (Z'')^3 - (3 Z (Z'')^2 - 3 Z^2 Z'')\big)| > \mu_{\alpha / 2}\right) \\
	        & = P\left(\frac{1}{\sqrt{15}}\trace(\tilde{Z}^3) > \mu_{\alpha / 2} - \frac{1}{\sqrt{15}} \trace((Z'')^3) + \frac{1}{\sqrt{15}} \trace\big(3 Z (Z'')^2 - 3 Z^2 Z''\big)\right) \\
	        & \quad + P\left(\frac{1}{\sqrt{15}}\trace(\tilde{Z}^3) < - \mu_{\alpha / 2} - \frac{1}{\sqrt{15}} \trace((Z'')^3) + \frac{1}{\sqrt{15}} \trace\big(3 Z (Z'')^2 - 3 Z^2 Z''\big)\right) \\
	        & \quad \to 1 \enskip .
    	\end{align*}
	
	Combing this with \eqref{trace4}, we have
	    \begin{align*}
	        P(|\hat \theta| > \mu_{\alpha / 2}) & = P(|\theta + o_p(1)| > \mu_{\alpha / 2}) \to 1 \enskip ,
	    \end{align*}
	which completes our proof.

\subsection{Proof of Theorem \ref{theorem theta hat}}\label{proof: theta hat}
    	
    It is not difficult to observe that for any $i \neq j$,
    	\begin{equation}\label{Z_hat and Z}    	
    	    \hat{Z}_{i j} = \frac{\sqrt{\frac{1}{m_1} \Sigma_{1, i j} + \frac{1}{m_2} \Sigma_{2, i j}}}{\sqrt{\frac{1}{m_1} \hat\Sigma_{1, i j} + \frac{1}{m_2} \hat\Sigma_{2, i j}}} Z_{i j}  \enskip .
    	\end{equation}
    Under the null hypothesis $H_0$, we have $\Sigma_1 = \Sigma_2$.
    Without loss of generality, assume $\hat\Sigma_{1, i j} \leq \hat\Sigma_{2, i j}$.
    For the numerator in \eqref{Z_hat and Z}, utilizing a Taylor expansion, we have	
    	\begin{align} 
    	    \sqrt{\frac{1}{m_1} \Sigma_{1, i j} + \frac{1}{m_2} \Sigma_{2, i j}} & = \sqrt{\frac{m_1 + m_2}{m_1 m_2}  \Sigma_{1, i j}} \nonumber\\
    	    & = \sqrt{\frac{m_1 + m_2}{m_1 m_2}} \Big( \sqrt{\hat\Sigma_{1, i j}} + O\big(\Sigma_{1, i j} - \hat{\Sigma}_{1, i j}\big)\Big) \nonumber\\ 
    	    & \leq \sqrt{\frac{\hat\Sigma_{1, i j}}{m_1 } + \frac{\hat\Sigma_{2, i j}}{m_1 }} + \sqrt{\frac{m_1 + m_2}{m_1 m_2}} O\big(\Sigma_{1, i j} - \hat{\Sigma}_{1, i j}\big) \nonumber\\
    	    & \leq \sqrt{\frac{\hat\Sigma_{1, i j}}{m_1 } + \frac{\hat\Sigma_{2, i j}}{m_1 }} + \sqrt{\frac{m_1 + m_2}{m_1 m_2}} o_p(1) \enskip , \label{approx1}
    	\end{align}
    where the last inequality comes from the  condition $\max_{i j}|\hat{\Sigma}_{u, i j} - \Sigma_{u, i j}| =  o_p(1)$.
    	
    Similarly, we have
    	\begin{equation}\label{approx2}
    	    \quad \sqrt{\frac{1}{m_1} \hat\Sigma_{1, i j} + \frac{1}{m_2} \hat\Sigma_{2, i j}} 
    	    \leq \sqrt{\frac{1}{m_1} \Sigma_{1, i j} + \frac{1}{m_2} \Sigma_{2, i j}} + \sqrt{\frac{m_1 + m_2}{m_1 m_2}} o_p(1) \enskip .
    	\end{equation}
    From \eqref{approx1} and \eqref{approx2}, 
    	\begin{equation}\label{approx 3}
    	    \quad \sqrt{  \frac{1}{m_1} \Sigma_{1, i j} + \frac{1}{m_2} \Sigma_{2, i j}  } 
    	    = \sqrt{  \frac{1}{m_1} \hat{\Sigma}_{1, i j} + \frac{1}{m_2} \hat{\Sigma}_{2, i j} } + \sqrt{ \frac{m_1 + m_2}{m_1 m_2} }  o_p(1) \enskip .
    	\end{equation}
    Combining \eqref{approx 3} with \eqref{Z_hat and Z} and noting that $Z_{ii} = \hat{Z}_{ii} = B_{ii}$, 
    we have
    	\begin{equation}\label{Z hat Z}
    	    \hat{Z} - B =   (J + H) \circ (Z - B) \enskip ,
    	\end{equation}
    where $H$ is an $n \times n$ matrix with entries $H_{i j} = o_p(1)$.
    	
    The remainder of the proof follows just the same as in Theorem \ref{theorem theta hat unweight}.

\subsection{Proof of \texorpdfstring{\eqref{error}}{SPE-AVG Error}} \label{proof: error}
Denote  $G_1$ and $G_2$ as two $n \times n$ matrices, with elements $G_{u, i j} = \mathrm{E}\big(\big({A_{u, i j}^{(k_u)}}\big)^2\big)$ for $u = 1, 2$, then
    \begin{align}
    & \quad | \hat{\Sigma}_{u, i j}^2 - \Sigma_{u, i j}^2 | \nonumber\\
	& = \left| \frac{\sum_{k_u = 1}^{m_u} \big({A_{u, i j}^{(k_u)}}\big)^2 - m_u \big(\bar{A}_{u, i j}\big)^2 }{m_u - 1} - G_{u, i j} + P_{u, i j}^2 \right| \nonumber \\
    & = \left| \frac{\sum_{k_u = 1}^{m_u} \big({A_{u, i j}^{(k_u)}}\big)^2}{m_u} - G_{u, i j} + \frac{\sum_{k_u = 1}^{m_u} \big({A_{u, i j}^{(k_u)}}\big)^2}{m_u (m_u -1)} 
    - \Bigg(\big(\bar{A}_{u, i j}\big)^2 -  P_{u, i j}^2 + \frac{\big(\bar{A}_{u, i j}\big)^2}{m_u - 1}\Bigg)\right|\nonumber \\
    & \leq \left| \frac{\sum_{k_u = 1}^{m_u} \big({A_{u, i j}^{(k_u)}}\big)^2}{m_u} - G_{u, i j}\right|  + \frac{\max\limits_{k_u = 1, 2, \dots, m_u} \big({A_{u, i j}^{(k_u)}}\big)^2}{m_u -1} \nonumber\\
    & \quad + |\bar{A}_{u, i j} -  P_{u, i j}| |\bar{A}_{u, i j} +  P_{u, i j}| + \frac{\big(\bar{A}_{u, i j}\big)^2}{m_u - 1} \nonumber\\
    & = O_p(m_u^{-1/2}) + \frac{\max\limits_{k_u = 1, 2, \dots, m_u} \big({A_{u, i j}^{(k_u)}}\big)^2}{m_u -1} 
    + O_p(m_u^{-1/2}) |\bar{A}_{u, i j} +  P_{u, i j}| + \frac{\big(\bar{A}_{u, i j}\big)^2}{m_u - 1} \nonumber\\
    & = O_p(m_u^{-1/2}),
    \end{align}
    where the second to last equality comes from Bernstein's inequality and the last equality is due to the boundedness of distributions $\{F_{u, i j}\}$ in our framework. It is noted that $\hat{\Sigma}_{u, i j}$ and $ \Sigma_{u, i j}$ are bounded, so $|\hat{\Sigma}_{u, i j} + \Sigma_{u, i j}|$ is also bounded. As a result, $\max\limits_{i, j}|\hat{\Sigma}_{u, i j} - \Sigma_{u, i j}| = O_p(m_u^{-1/2})$.

\end{document}